\documentclass[
    aps,
    prl,
    reprint,
    superscriptaddress,
    longbibliography,
    floatfix
]{revtex4-2}

\usepackage{amsmath,amssymb}
\usepackage{graphicx}
\usepackage{array}[=2016-10-06]
\usepackage{siunitx}
\usepackage[hidelinks]{hyperref}

\graphicspath{{fig/}}

\begin{document}

\title{All-Dielectric Tight-Binding Photonics via a Band-Gap Vacuum}
\author{Aodong Li}
\thanks{These authors contributed equally to this work.}
\author{Jing Li}
\thanks{These authors contributed equally to this work.}
\author{Tao Xiao}
\author{Biye Xie}
\email{xiebiye@cuhk.edu.cn}
\affiliation{School of Science and Engineering, The Chinese University of Hong Kong (Shenzhen), Longgang, Shenzhen, Guangdong 518172, China}

\begin{abstract}
Tight-binding photonics bridges condensed-matter physics and optics, providing a powerful framework for engineering band structures and topological phases with light. For photons, however, the vacuum is gapless, so resonators in air carry radiative tails that produce spurious long-range couplings. Here we embed resonators in a photonic band-gap vacuum (BgV), a photonic crystal whose band gap plays for photons the role vacuum plays for bound electrons. The BgV localizes each resonance exponentially, so designed short-range couplings define the lattice Hamiltonian. We thereby experimentally realize a faithful Benalcazar--Bernevig--Hughes Hamiltonian in an all-dielectric photonic crystal, resolving corner resonances and their orbitals in microwave measurements. Unlike the vacuum surrounding electrons, the BgV is itself designable, controlling relative couplings and closed-loop interference through its band dispersion and Bloch modes. Full-wave calculations and extracted models show that uniformly modifying the BgV, with all resonators fixed, switches the quantized polarization of a two-dimensional Su--Schrieffer--Heeger lattice and converts a zero-flux lattice into a $\pi$-flux one. We also numerically realize three-dimensional pyrochlore and hyperhoneycomb lattices with space groups $R\bar{3}m$ and $C2/c$, the latter a nodal-loop semimetal with topological drumhead surface states. These results establish the BgV as a designable vacuum for all-dielectric tight-binding photonics in two and three dimensions.
\end{abstract}

\maketitle

\textit{Introduction.---}
Photonic crystals originated from the insight of Yablonovitch and John that a photonic band gap can reshape the electromagnetic vacuum itself, inhibiting spontaneous emission and localizing light within an engineered dielectric environment \cite{Yablonovitch1987Inhibited,John1987Strong}. They are now central to nanophotonics and integrated optics \cite{Notomi2008CoupledNanocavities,Han2019TopologicalCavity,Busch2007Periodic}. More recently, tight-binding photonics \cite{Li2026TightBinding} has attracted wide interest by casting localized resonances and their couplings as a lattice Hamiltonian, bringing condensed-matter concepts such as topological boundary states to optics \cite{Lidorikis1998TightBinding,Ozawa2019Topological,Zhang2023SecondWave,Leykam2026Limitations,Hasan2010Colloquium,Qi2011Topological,Su1979Solitons,Haldane1988Model,KaneMele2005QuantumSpinHall,Bernevig2006QuantumSpinHall,Benalcazar2017Quantized,Bradlyn2017Topological}. The vacuum, however, plays fundamentally different roles for electrons and photons. For a bound electron it is an insulating barrier into which orbitals decay exponentially, whereas for a photon it is gapless, so dielectric resonators in air carry radiative tails that produce unwanted long-range couplings \cite{Lidorikis1998TightBinding,Li2024Disentangled}.

Tight-binding photonics has therefore relied on structures that avoid free-space radiation. Waveguide arrays \cite{Schulz2022PhotonicQuadrupole,Rechtsman2013Photonic,Maczewsky2020Nonlinearity} use one coordinate for evolution, limiting lattices to two dimensions. Coupled-resonator optical waveguides and cavity arrays realize tight-binding models via localized-mode coupling \cite{Yariv1999,Bayindir2000TightBinding,Notomi2008CoupledNanocavities,Dossou2009,Han2019TopologicalCavity}, but are primarily chains or planar arrays. Confined-Mie-resonance photonic crystals suppress radiative tails with metallic inclusions \cite{Li2024Disentangled}, enabling three-dimensional higher-order topological insulators \cite{Wang2025ThreeDimensional}, synthetic $\pi$ flux \cite{Huang2025SyntheticFlux,Li2026GaugeDelocalization}, and three-dimensional Dirac-vortex modes \cite{Yan2025DiracVortex}. Metallic absorption, however, grows with frequency and hinders their extension to optical frequencies. Two- and three-dimensional all-dielectric photonic crystals could overcome these limitations for on-chip photonics, semiconductor lasers, and quantum light--matter interfaces \cite{Notomi2008CoupledNanocavities,Han2019TopologicalCavity,Douglas2015QuantumManyBody,Hood2016BandEdge,Guo2026DefectEvolved}.

Here we embed resonators in a photonic band-gap vacuum (BgV), a photonic crystal whose band gap plays for photons the role vacuum plays for bound electrons. The BgV serves two complementary functions, confinement and designability. With no propagating states in the gap, each embedded resonance acquires an exponentially decaying tail, suppressing remote couplings while retaining the designed short-range ones. The lattice Hamiltonian is then set by site placement and orbital choice, as for atoms in a crystal. We test this with the Benalcazar--Bernevig--Hughes (BBH) model of quantized multipole topology \cite{Benalcazar2017Quantized}, previously realized in mechanical \cite{SerraGarcia2018Phononic}, circuit \cite{Peterson2018MicrowaveQuadrupole,Imhof2018Topolectrical}, ring-resonator \cite{Mittal2019Quadrupole}, and waveguide \cite{Schulz2022PhotonicQuadrupole} platforms. Placing air-cavity and $p$-orbital resonance modes in a dielectric-rod BgV, we implement the BBH Hamiltonian with $\pi$ flux per plaquette in an all-dielectric photonic crystal slab and observe corner resonances in microwave measurements.

Unlike electronic vacuum, the engineered BgV offers a second design level through its band dispersion and Bloch modes, which determine evanescent-tail decay and phase \cite{Dossou2009}. Uniform BgV changes tune inequivalent links differently at fixed resonators, reshaping closed-loop interference and complementing local tuning \cite{Kee2003,Haddadi2014,Zheng2026Tailored}. Full-wave calculations and extracted models show that BgV tuning switches quantized polarization in a two-dimensional Su--Schrieffer--Heeger (SSH) lattice, and converts zero to $\pi$ flux in a separate lattice.

In three dimensions, besides confinement of resonance modes, a further challenge arises as the electromagnetic waves are generally vector fields instead of scalar fields and cannot be simply decoupled into transverse-electric (TE) and transverse-magnetic (TM) polarizations. Here we consider a 3D layered photonic crystal with a complete band gap as the BgV \cite{Johnson2000Layered,Povinelli2001Emulation}. Moreover, this unique photonic crystal supports TM-like modes in the engineered resonant cavities in the rod layers, allowing us to design 3D photonic pyrochlore \cite{Ezawa2018Breathing} and hyperhoneycomb (HHC) \cite{Ezawa2016LoopNodal} lattices with distinct symmetries and topology. Our result of achieving all-dielectric tight-binding photonics based on the BgV opens a route toward novel photonic devices at optical frequencies.

\begin{figure}[!t]
    \centering
    \includegraphics[width=\columnwidth]{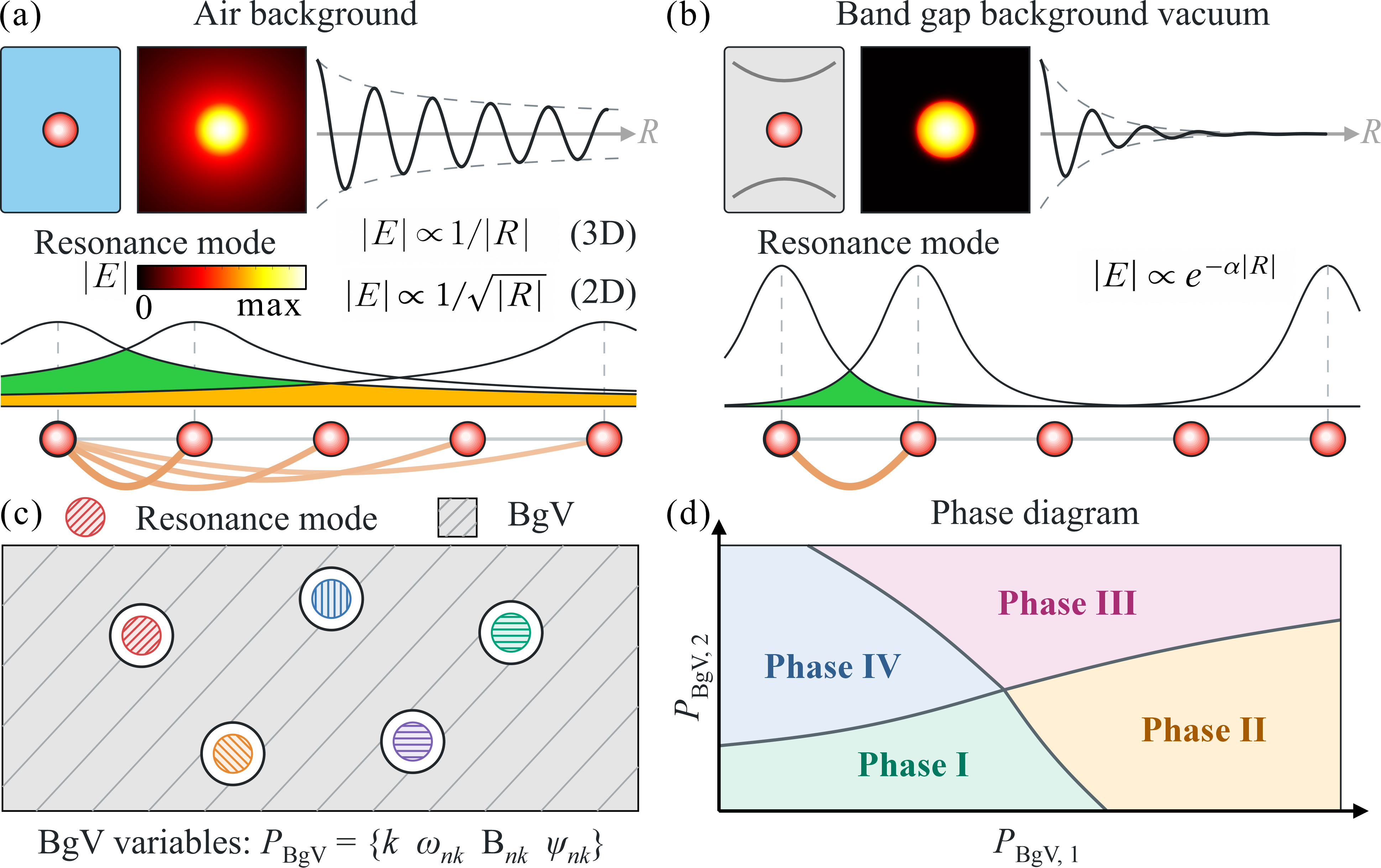}
    \caption{Resonance-mode and BgV controls.
    (a) Radiative far-field tails decay as $1/|R|$ (3D) and $1/\sqrt{|R|}$ (2D), overlapping neighboring (green) and remote (yellow) sites. (b) Exponential in-gap field decay suppresses remote overlap.
    (c) Colored interiors denote resonance-mode degrees of freedom, including orbital type and angular momentum, independent of BgV tuning. Black circles mark fixed resonator geometries and positions, with BgV-induced field reshaping retaining resonator character. (d) Conceptual phase
    diagram of the BgV control space $P_{\mathrm{BgV}}$.}
    \label{fig:fig1}
\end{figure}

\textit{Band-gap vacuum.---}
To identify how the BgV controls coupling, we compare mode tails in air and in a BgV [Figs.~\ref{fig:fig1}(a,b)]. Radiative tails in air produce long-range overlap, whereas in-gap tails have exponentially decaying envelopes, enabling a short-range model when omitted couplings are small on the relevant spectral scales.

For a lossless, nondispersive, nonmagnetic periodic BgV with scalar
relative permittivity $\varepsilon_{\mathrm b}( R)>0$ in spatial
dimension $D=2$ or $3$, the in-gap Green tensor is
\cite{Dossou2009,Dossou2008Shallow,Busch2007Periodic,Silveirinha2021}
\begin{equation}
\scalebox{0.95}{$\displaystyle
 G_{\mathrm{BgV}}( R,\! R';\omega)
\!=\!\frac{c^2}{(2\pi)^D}\!\sum_n\!\int_{\mathrm{BZ}}\!
\frac{\psi_{n k}( R)\!\otimes\!
\psi_{n k}^{*}( R')}{\omega^2-\omega_{n k}^{2}}\,d^D k.$}
\label{eq:bg-green}
\end{equation}
Here $\psi_{n k}$ and $\omega_{n k}$ are bulk electric-field Bloch modes and angular eigenfrequencies, with band index $n$ and wave vector $k$ (inverse length) in the first Brillouin zone (BZ). $ R, R'$ are observation and source positions, $c$ is the
vacuum speed of light, and $\otimes$ denotes a dyadic product. The in-gap angular frequency is $\omega=2\pi f$, with ordinary frequency $f$.

An isolated in-gap resonance field $ E_i$ centered at $ R_i$, with angular
frequency $\omega_i$ and relative permittivity
$\varepsilon_i=\varepsilon_{\mathrm b}+\Delta\varepsilon_i$, satisfies
\begin{equation}
\scalebox{0.95}{$\displaystyle
 E_i( R)\!=\!-\frac{\omega_i^2}{c^2}\!\int_{\mathcal V_i}\!
\Delta\varepsilon_i( R')
 G_{\mathrm{BgV}}( R,\! R';\omega_i)
 E_i( R')\,d^D\!R'.$}
\label{eq:localized-field}
\end{equation}
Here $\Delta\varepsilon_i$ is the dielectric contrast in resonator region $\mathcal V_i$.
For weakly coupled fields $E_i,E_j$ with $\omega_i\simeq\omega_j\simeq\omega_0$, the leading coupling in ordinary-frequency units is \cite{Yariv1999}
\begin{equation}
t_{ij}\simeq-\frac{\omega_0}{4\pi\sqrt{ Z_i Z_j}}
\int d^D R\,\Delta\varepsilon_i( R)
 E_i^*( R)\!\cdot\! E_j( R).
\label{eq:coupling-overlap}
\end{equation}
The normalization integrals are $Z_i=\int d^D R\,\varepsilon_i|E_i|^2$ and similarly $Z_j$.
Equation~(\ref{eq:coupling-overlap}) agrees with couplings extracted from two-cavity frequency splitting (Supplemental Material, Note S1 \cite{SupplementalMaterial}).

Equations~(\ref{eq:bg-green})--(\ref{eq:coupling-overlap}) connect orbital structure and site placement to the BgV control space $P_{\mathrm{BgV}}=(k,\omega_{n k},B_{n k},\psi_{n k})$, where $B_{n k}$ is the Hessian matrix of $\omega_{n k}$ with respect to $k$, describing the local band curvature. Near regular band edges, curvature and detuning determine the directional decay coefficient $\alpha$ \cite{Dossou2009}, while Bloch fields shape spatial phase. Changes of the BgV modify evanescent fields, reshaping couplings at fixed resonator geometries, positions, and orbital types [Fig.~\ref{fig:fig1}(c)], and can drive topological transitions [Fig.~\ref{fig:fig1}(d)].

\begin{figure}[!t]
    \centering
    \includegraphics[width=\columnwidth]{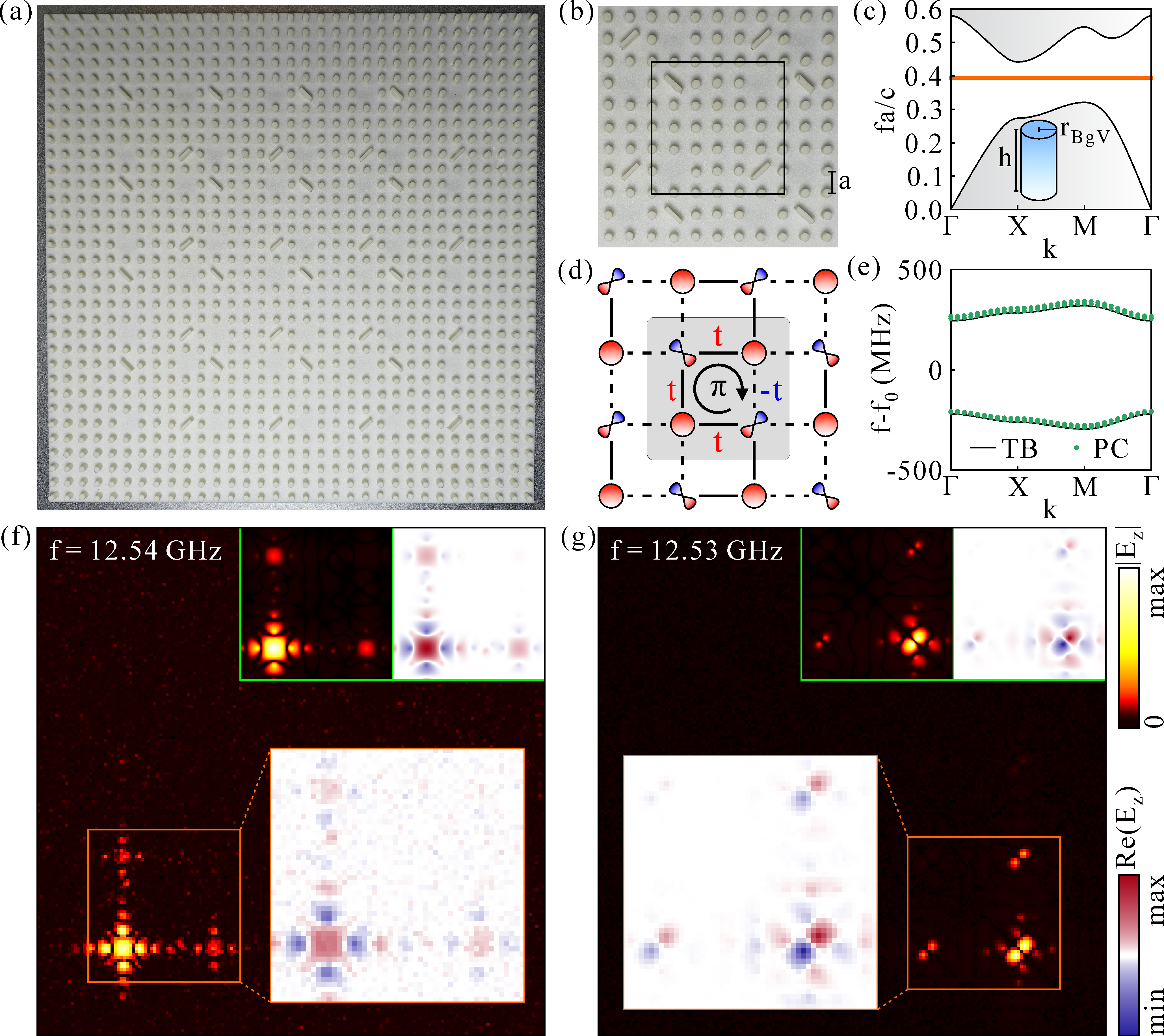}
    \caption{Experimental BBH photonic lattice and corner response.
    (a) Fabricated $4\times4$-cell alumina sample. (b) Enlarged region, with a unit cell outlined in black. (c) Simulated 2D TM BgV bands and resonance mode (orange), $r_{\mathrm{BgV}}=0.2a$.
    (d) Four-site BBH lattice, with solid/dashed bonds denoting positive/negative couplings. (e) Tight-binding (TB, black)
    and simulated photonic-crystal (PC, green) bands
    relative to the isolated resonance frequency $f_0$.
    (f,g) Measured normalized probe-response amplitudes at \SI{12.54}{GHz} and \SI{12.53}{GHz}. Orange-framed insets show measured local $\operatorname{Re}(E_z)$. Green-framed upper-right insets show simulated $|E_z|$ (left) and $\operatorname{Re}(E_z)$ (right).}
    \label{fig:fig2}
\end{figure}

\textit{All-dielectric photonic quadrupole insulator.---}
We experimentally realize BBH couplings through positional and orbital control in a fixed dielectric-rod BgV [Figs.~\ref{fig:fig2}(a,b)] with $\varepsilon_r=9$ (all two-dimensional models), lattice constant $a=\SI{10}{mm}$, rod radius $r_{\mathrm{BgV}}=0.2a$, and height $h=0.8a$. Removing one rod creates an air-cavity mode in the TM gap, with $E_z$ along the rod-height direction [Fig.~\ref{fig:fig2}(c)].

Combining $s$-orbital modes with frequency-matched $p$-orbital modes reverses selected couplings through $p$-mode odd parity \cite{Li2024Disentangled,Huang2025SyntheticFlux} and, with site placement, produces the BBH $\pi$-flux pattern [Fig.~\ref{fig:fig2}(d)].
Individual coupling signs depend on the resonance-mode phase convention, but the oriented-loop flux $\Phi_{\mathcal C}=\arg\prod_{(ij)\in\mathcal C}t_{ij}$ is gauge invariant, equaling $0$ or $\pi$ for nonzero real reciprocal couplings.

We compare full-wave bands with the four-site BBH Hamiltonian \cite{Benalcazar2017Quantized} (End Matter), whose eigenvalues are $f-f_0$, with $f_0=\omega_0/(2\pi)$. For the two-dimensional reference model,
the cavities are arranged along the BgV lattice vectors in the $x$ and $y$
directions, with coupling magnitudes that decay exponentially with separation
(Supplemental Material, Note S2 \cite{SupplementalMaterial}).
The intercell coupling magnitude exceeds the intracell value.
Full-wave bands reproduce the gapped BBH dispersion [Fig.~\ref{fig:fig2}(e)]. This topological configuration supports in-gap modes at all four corners under open boundaries.
We probe corner orbitals in a fabricated $4\times4$-cell alumina sample [Figs.~\ref{fig:fig2}(a,b)] (experimental details in Supplemental Material, Note S3 \cite{SupplementalMaterial}). Measured responses localize at the lower-left and lower-right corners at \SI{12.54}{GHz} and \SI{12.53}{GHz}, respectively [Figs.~\ref{fig:fig2}(f,g)]. The normalized complex probe response, proportional to $E_z$, resolves orbital profiles including the lower-right $p$ mode, matching simulated corner localization and orbitals [upper-right insets in Figs.~\ref{fig:fig2}(f,g)].
Within the same BgV gap, field measurements show intersite propagation at neighboring BBH-array bulk-band frequencies but corner localization at \SI{12.54}{GHz}, excluding background-gap confinement alone as its cause (Supplemental Material, Note S3, Fig.~S11 \cite{SupplementalMaterial}).
At the measurement frequencies, the metal plates approximate perfect
electric conductors and confine the field vertically to reproduce the
two-dimensional TM setting in the three-dimensional experiment.
Full-wave simulations retain the characteristic BBH dispersion for taller rods without metal plates, although finite radiative linewidths remain (Supplemental Material, Note S3, Fig.~S12 \cite{SupplementalMaterial}).
\nocite{Waks2005CoupledMode}

\begin{figure}[!t]
    \centering
    \includegraphics[width=\columnwidth]{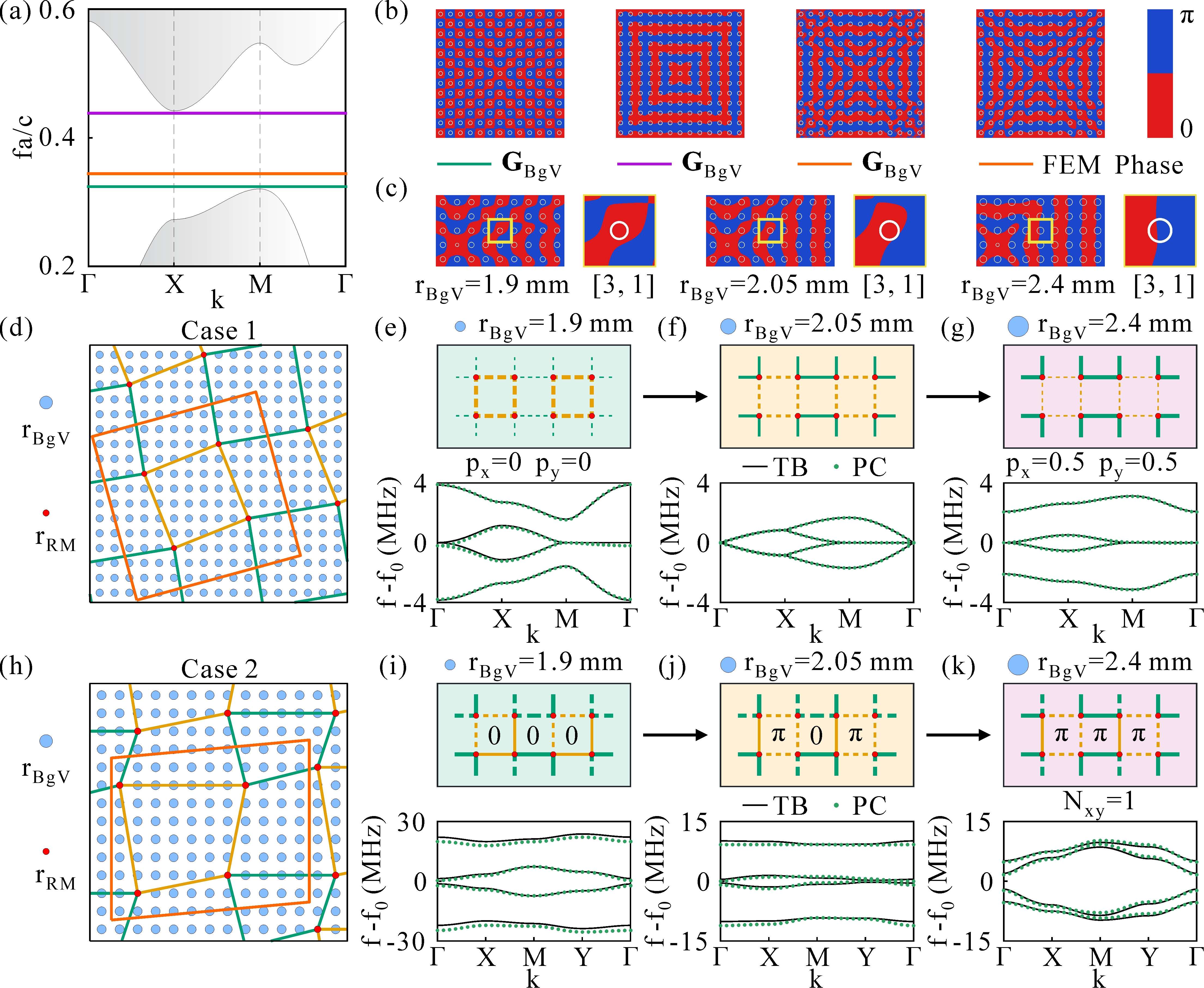}
    \caption{BgV control of tails and couplings.
    (a) BgV bands and three in-gap frequencies.
    (b) Point-source Green-function phases from the first ten BgV bands in
    Eq.~(\ref{eq:bg-green}): near $M$ (green), near $X$ (purple),
    and inside the gap (orange), followed by the FEM localized-field phase
    at the orange frequency. (c) Localized phases at $r_{\mathrm{BgV}}=1.9$,
    $2.05$, and $\SI{2.4}{mm}$, with insets enlarging the tail-field phase
    at the rod displaced by $[3,1]a$ from the mode center.
    Red/blue denote field phases of $0/\pi$ in (b,c).
    (d) Lattice for BgV-controlled topological bulk polarization, with fixed resonator-rod radius $r_{\mathrm{RM}}=\SI{1.2}{mm}$.
    Yellow/green links are
    intracell/intercell, and orange outlines mark unit cells.
    (e--g) SSH patterns and bands for different BgV-rod radii.
    (h) Lattice for BgV-controlled loop flux at the same fixed resonator-rod radius.
    (i--k) Chiral-lattice patterns and bands for different BgV-rod radii,
    with different lattice loop fluxes. Black/green denote TB/PC bands relative
    to each BgV's localized-mode frequency $f_0$.}
    \label{fig:fig3}
\end{figure}

\begin{figure*}[!t]
    \centering
    \includegraphics[width=\textwidth]{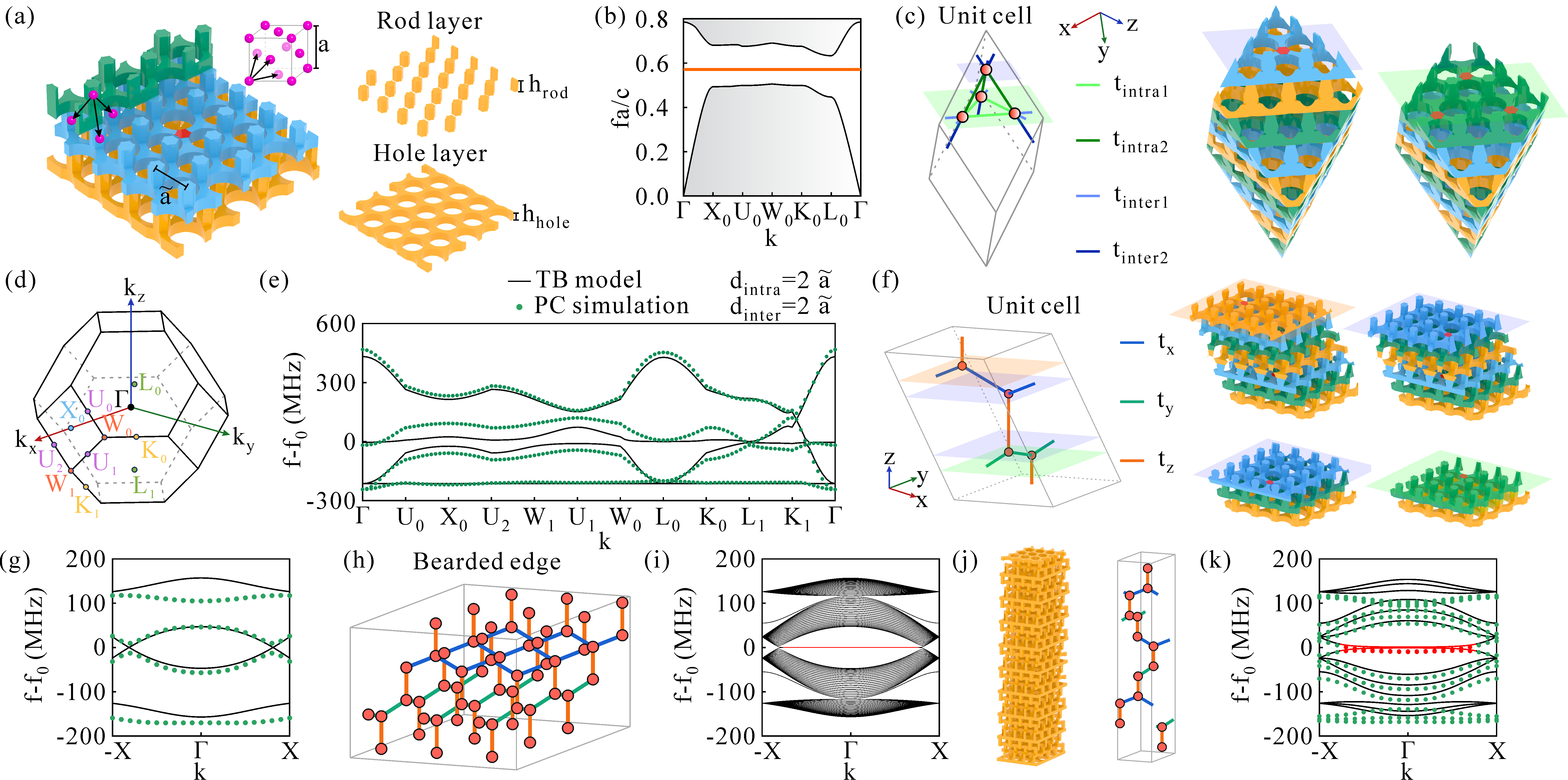}
    \caption{Three-dimensional lattices in an fcc BgV.
    (a) Stacked crystal, rod/hole layers, and fcc cell (inset).
    (b) BgV bands and resonance mode (orange).
    (c) Pyrochlore cell and implementation with equal intracell and intercell
    site separations $d_{\mathrm{intra}}=d_{\mathrm{inter}}=2\tilde a$.
    Light/dark colors distinguish lateral/oblique links, and green/blue
    distinguish intracell/intercell links. (d) Brillouin zone of the pyrochlore cell.
    (e) Bands of the gapless pyrochlore configuration.
    (f) HHC cell and implementation, with blue, green, and orange bonds denoting $t_x,t_y,t_z$.
    (g) HHC bulk bands. (h) HHC lattice with a bearded termination. (i) Model
    spectrum for 20 periods along $z$, with drumhead surface states in red.
    (j) Three-period slab and model. (k) Three-period spectra, where red dots
    denote the drumhead surface states. In (e,g,k), black lines and
    green/red dots denote tight-binding and full-wave results.}
    \label{fig:fig4}
\end{figure*}

\textit{Uniform BgV control of tight-binding photonics.---}
To examine BgV designability numerically, we evaluate the scalar TM Green function $G_{\mathrm{BgV}}(R,0;\omega)$ for a point source at a BgV-rod center and three in-gap frequencies at $r_{\mathrm{BgV}}=\SI{2}{mm}$, retaining the first ten bands in Eq.~(\ref{eq:bg-green}) [Figs.~\ref{fig:fig3}(a,b)]. Near the lower ($M$) or upper ($X$) edge, phases resemble the corresponding band-edge Bloch fields, whereas both bordering bands dominate deeper in the gap. At the orange-marked frequency, finite-element-method (FEM) localized-field phases follow the main Green-function phase domains. Most discrepancies occur near outer corners, where weak fields make phase sensitive to small field variations.

Varying $r_{\mathrm{BgV}}$ at fixed local-rod radius $r_{\mathrm{RM}}=\SI{1.2}{mm}$ and mode center shifts the in-gap resonance frequency and reshapes its evanescent-tail phase (Supplemental Material, Note S1 \cite{SupplementalMaterial}).
At displacement $[3,1]a$ from the mode center, the tail changes from predominantly $0$ (red) to $\pi$ (blue) phase as $r_{\mathrm{BgV}}$ increases from \SI{1.9}{mm} to \SI{2.4}{mm} [Fig.~\ref{fig:fig3}(c)]. With a fixed
resonance-mode phase convention, this redistribution reverses the signed overlap with a receiving mode in
Eq.~(\ref{eq:coupling-overlap}), consistent with the coupling reversal
at this displacement (Supplemental Material, Note S4
\cite{SupplementalMaterial}). Different band-edge Bloch contributions carry distinct spatial phases, weighted by spectral detuning and local-orbital projections, so their interference can cancel overlaps and reverse selected coupling signs. The spatial phase depends on displacement,
allowing a uniform BgV change to tune distinct bonds differently
(Supplemental Material, Note S1 \cite{SupplementalMaterial}).

By reshaping evanescent tails between fixed resonators, BgV tuning alone changes the topological phase of the SSH array [Fig.~\ref{fig:fig3}(d)] without changing resonator geometries. Intracell dominance at $r_{\mathrm{BgV}}=\SI{1.9}{mm}$ gives $(p_x,p_y)=(0,0)$ [Fig.~\ref{fig:fig3}(e)]. The lowest-band direct gap nearly closes at \SI{2.05}{mm} and reopens with intercell dominance at \SI{2.4}{mm}, where $(p_x,p_y)=(1/2,1/2)$ [Figs.~\ref{fig:fig3}(f,g)]. These inversion-quantized polarizations belong to the extracted model's isolated lowest band, modulo one in target-translation units. Although the intercell link is longer, its coupling magnitude overtakes the intracell value near $r_{\mathrm{BgV}}=\SI{2.05}{mm}$. Model bands closely match full-wave results (coupling curves in Supplemental Material, Fig.~S13 \cite{SupplementalMaterial}).

Beyond relative coupling magnitudes, BgV tuning can control closed-loop flux through coupling-sign reversals. We illustrate this with a generalized four-site chiral lattice [Fig.~\ref{fig:fig3}(h)] \cite{Li2018CornerConfigurations}. This model evolves from uniform zero flux at \SI{1.9}{mm}, through $0/\pi$ flux alternating along $x$ at \SI{2.05}{mm}, to uniform $\pi$ flux at \SI{2.4}{mm} [Figs.~\ref{fig:fig3}(i--k)], reproducing the main full-wave band features along the plotted path. At the uniform-$\pi$ endpoint, the principal-link Hamiltonian forms a chiral BBH-like model with multipole chiral number $N_{xy}=1$ \cite{Benalcazar2022Chiral}. Figure~S14 in Supplemental Material, Note S4 \cite{SupplementalMaterial}, shows the full flux sequence. The SSH transition and flux conversion demonstrate changes in the coupling hierarchy and closed-loop interference, respectively, at fixed resonator geometries and positions while following the same orbital branch. BgV tuning can also suppress a selected coupling and, with onsite-frequency compensation, produce a nearly flat middle band in a rhombic chain (Supplemental Material, Note S4, Fig.~S15 \cite{SupplementalMaterial}).

\textit{3D all-dielectric tight-binding photonics.---}
Our three-dimensional single-orbital lattice requires a complete gap, isolated vector resonances at each site, and controlled directional overlaps. To meet these requirements, we use
an all-dielectric face-centered-cubic (fcc) crystal \cite{Johnson2000Layered,Povinelli2001Emulation}
stacked along [111] [Fig.~\ref{fig:fig4}(a)]. The dielectric components have
relative permittivity $\varepsilon_r=12$ in this fcc BgV.
Each period combines triangular-lattice rod and air-hole layers with $h_{\mathrm{rod}}+h_{\mathrm{hole}}=a/\sqrt{3}$ and $h_{\mathrm{rod}}/h_{\mathrm{hole}}=1.57$. Here $a=\SI{10}{mm}$ and
the minimum fcc-site spacing is $\tilde a=a/\sqrt{2}$.
Green, blue, and yellow mark three laterally offset layers.  The structure has the complete
gap in Fig.~\ref{fig:fig4}(b).  Removing selected rods creates localized modes
whose evanescent coupling forms the target lattice.

Embedding localized modes in the fcc BgV yields a pyrochlore model \cite{Ezawa2018Breathing} [Fig.~\ref{fig:fig4}(c)]. At equal separations $2\tilde a$, its Hamiltonian (End Matter) captures the main full-wave dispersion and bandwidth [Figs.~\ref{fig:fig4}(d,e)]. Lateral and oblique bonds are inequivalent in the layered BgV, giving this Hamiltonian trigonal symmetry $R\bar{3}m$ (No.~166), a subgroup of ideal pyrochlore symmetry $Fd\bar{3}m$ (No.~227).

Another fcc-site subset forms the three-coordinated $[xy]$ HHC lattice \cite{Ezawa2016LoopNodal}, with four full-wave bands inside the BgV gap [Figs.~\ref{fig:fig4}(f,g)]. Its Hamiltonian has a nodal loop in a tilted momentum plane (End Matter and Supplemental Material, Note S5 \cite{SupplementalMaterial}). For $t_x=t_y$, the model has monoclinic symmetry $C2/c$ (No.~15) \cite{Corticelli2022SpinSpace}, with the nodal loop protected by sublattice and $PT$ symmetries \cite{Wawrzik2018Honeycomb}.

A 20-period tight-binding slab with bearded $z$ termination [Fig.~\ref{fig:fig4}(h)] hosts topological drumhead surface states inside the nodal-loop projection onto the surface Brillouin zone [Fig.~\ref{fig:fig4}(i)] \cite{Wawrzik2018Honeycomb}. Memory limits full-wave HHC simulations to three $z$ periods with transverse periodic boundaries [Fig.~\ref{fig:fig4}(j)]. The full-wave spectrum approximately follows the model, including drumhead states [Fig.~\ref{fig:fig4}(k)]. The full-wave field of the $\Gamma$-point drumhead state [red dot in Fig.~\ref{fig:fig4}(k)] localizes at the bearded boundaries (Supplemental Material, Note S5 \cite{SupplementalMaterial}). Site selection sets connectivity, while the BgV sets coupling strengths and anisotropy.

\begingroup
\clubpenalty=10000
\widowpenalty=10000
\textit{Discussion and outlook.---}
We have demonstrated all-dielectric tight-binding photonics using the BgV. This approach combines exponential confinement with multiparameter control of the Hamiltonian. Two limitations remain. First, BgV-induced changes in resonance modes may not be fully captured by the tight-binding approximation, leading to small quantitative discrepancies between the photonic-crystal and tight-binding bands (Supplemental Material, Note S4, Fig.~S15 \cite{SupplementalMaterial}). Second, the vacuum analogy becomes exact only in the continuum limit. Approaching this limit requires a small background period relative to resonator spacing. Increasing the spacing, however, typically weakens coupling exponentially and narrows the bandwidth. Locally resonant dielectrics with subwavelength gaps could alleviate this trade-off by reducing the background scale at fixed resonator spacing \cite{Lemoult2013WavePropagation}.

Active, gyromagnetic, and nonlinear BgVs offer further possibilities. Patterned gain and loss could generate complex reciprocal couplings, allowing one background parameter to drive detuned resonances toward exceptional points or selectively amplify collective modes through their overlap with the background \cite{Miri2019ExceptionalPoints}. Self-biased ferrites with stable remanent magnetization could break time-reversal symmetry without sustained external fields \cite{Yang2023SelfBiased}, introducing complex coupling phases and nonuniform fluxes such as Harper--Hofstadter patterns \cite{Wang2009Unidirectional,Hofstadter1976EnergyLevels}. A Kerr BgV could make coupling ratios intensity dependent, enabling all-optical control and self-induced topological transitions \cite{Maczewsky2020Nonlinearity}. The vacuum itself thus becomes a material degree of freedom for controlling collective photonic phases, extending band-gap engineering beyond spontaneous emission \cite{Yablonovitch1987Inhibited,John1990QuantumElectrodynamics}.

\par\endgroup

\vspace{0.6\baselineskip}
\begin{acknowledgments}
B.~X. acknowledges support from the National Natural Science Foundation of
China (Grant Nos.~62475225 and 12404187), the National Key R\&D Program of
China (Grant Nos.~2023YFA1407700 and 2025YFA1412300), the Guangdong Basic and
Applied Basic Research Foundation (Grant No.~2024A1515012031), the Shenzhen
Science and Technology Innovation Commission (Grant No.~JCYJ20240813113619025),
and the Guangdong Provincial Quantum Science Strategic Initiative
(Grant No.~GDZX2604007).
A.~L. acknowledges support from the National Natural Science Foundation of
China (Grant No.~12347161) and the China Postdoctoral Science Foundation
(Grant Nos.~2024M753077 and 2024T170872).

\end{acknowledgments}

\begingroup
\clubpenalty=10000
\textit{Data availability.---}
The data and code supporting this study are available from the corresponding
author.
\par\endgroup

\bibliography{Ref/references}

\begin{thebibliography}{57}%
\makeatletter
\providecommand \@ifxundefined [1]{%
 \@ifx{#1\undefined}
}%
\providecommand \@ifnum [1]{%
 \ifnum #1\expandafter \@firstoftwo
 \else \expandafter \@secondoftwo
 \fi
}%
\providecommand \@ifx [1]{%
 \ifx #1\expandafter \@firstoftwo
 \else \expandafter \@secondoftwo
 \fi
}%
\providecommand \natexlab [1]{#1}%
\providecommand \enquote  [1]{``#1''}%
\providecommand \bibnamefont  [1]{#1}%
\providecommand \bibfnamefont [1]{#1}%
\providecommand \citenamefont [1]{#1}%
\providecommand \href@noop [0]{\@secondoftwo}%
\providecommand \href [0]{\begingroup \@sanitize@url \@href}%
\providecommand \@href[1]{\@@startlink{#1}\@@href}%
\providecommand \@@href[1]{\endgroup#1\@@endlink}%
\providecommand \@sanitize@url [0]{\catcode `\\12\catcode `\$12\catcode
  `\&12\catcode `\#12\catcode `\^12\catcode `\_12\catcode `\%12\relax}%
\providecommand \@@startlink[1]{}%
\providecommand \@@endlink[0]{}%
\providecommand \url  [0]{\begingroup\@sanitize@url \@url }%
\providecommand \@url [1]{\endgroup\@href {#1}{\urlprefix }}%
\providecommand \urlprefix  [0]{URL }%
\providecommand \Eprint [0]{\href }%
\providecommand \doibase [0]{https://doi.org/}%
\providecommand \selectlanguage [0]{\@gobble}%
\providecommand \bibinfo  [0]{\@secondoftwo}%
\providecommand \bibfield  [0]{\@secondoftwo}%
\providecommand \translation [1]{[#1]}%
\providecommand \BibitemOpen [0]{}%
\providecommand \bibitemStop [0]{}%
\providecommand \bibitemNoStop [0]{.\EOS\space}%
\providecommand \EOS [0]{\spacefactor3000\relax}%
\providecommand \BibitemShut  [1]{\csname bibitem#1\endcsname}%
\let\auto@bib@innerbib\@empty
\bibitem [{\citenamefont {Yablonovitch}(1987)}]{Yablonovitch1987Inhibited}%
  \BibitemOpen
  \bibfield  {author} {\bibinfo {author} {\bibfnamefont {E.}~\bibnamefont
  {Yablonovitch}},\ }\bibfield  {title} {\bibinfo {title} {Inhibited
  spontaneous emission in solid-state physics and electronics},\ }\href
  {https://doi.org/10.1103/PhysRevLett.58.2059} {\bibfield  {journal} {\bibinfo
   {journal} {Phys. Rev. Lett.}\ }\textbf {\bibinfo {volume} {58}},\ \bibinfo
  {pages} {2059} (\bibinfo {year} {1987})}\BibitemShut {NoStop}%
\bibitem [{\citenamefont {John}(1987)}]{John1987Strong}%
  \BibitemOpen
  \bibfield  {author} {\bibinfo {author} {\bibfnamefont {S.}~\bibnamefont
  {John}},\ }\bibfield  {title} {\bibinfo {title} {Strong localization of
  photons in certain disordered dielectric superlattices},\ }\href
  {https://doi.org/10.1103/PhysRevLett.58.2486} {\bibfield  {journal} {\bibinfo
   {journal} {Phys. Rev. Lett.}\ }\textbf {\bibinfo {volume} {58}},\ \bibinfo
  {pages} {2486} (\bibinfo {year} {1987})}\BibitemShut {NoStop}%
\bibitem [{\citenamefont {Notomi}\ \emph {et~al.}(2008)\citenamefont {Notomi},
  \citenamefont {Kuramochi},\ and\ \citenamefont
  {Tanabe}}]{Notomi2008CoupledNanocavities}%
  \BibitemOpen
  \bibfield  {author} {\bibinfo {author} {\bibfnamefont {M.}~\bibnamefont
  {Notomi}}, \bibinfo {author} {\bibfnamefont {E.}~\bibnamefont {Kuramochi}},\
  and\ \bibinfo {author} {\bibfnamefont {T.}~\bibnamefont {Tanabe}},\
  }\bibfield  {title} {\bibinfo {title} {Large-scale arrays of ultrahigh-{Q}
  coupled nanocavities},\ }\href {https://doi.org/10.1038/nphoton.2008.226}
  {\bibfield  {journal} {\bibinfo  {journal} {Nat. Photonics}\ }\textbf
  {\bibinfo {volume} {2}},\ \bibinfo {pages} {741} (\bibinfo {year}
  {2008})}\BibitemShut {NoStop}%
\bibitem [{\citenamefont {Han}\ \emph {et~al.}(2019)\citenamefont {Han},
  \citenamefont {Lee}, \citenamefont {Callard}, \citenamefont {Seassal},\ and\
  \citenamefont {Jeon}}]{Han2019TopologicalCavity}%
  \BibitemOpen
  \bibfield  {author} {\bibinfo {author} {\bibfnamefont {C.}~\bibnamefont
  {Han}}, \bibinfo {author} {\bibfnamefont {M.}~\bibnamefont {Lee}}, \bibinfo
  {author} {\bibfnamefont {S.}~\bibnamefont {Callard}}, \bibinfo {author}
  {\bibfnamefont {C.}~\bibnamefont {Seassal}},\ and\ \bibinfo {author}
  {\bibfnamefont {H.}~\bibnamefont {Jeon}},\ }\bibfield  {title} {\bibinfo
  {title} {Lasing at topological edge states in a photonic crystal {L3}
  nanocavity dimer array},\ }\href {https://doi.org/10.1038/s41377-019-0149-7}
  {\bibfield  {journal} {\bibinfo  {journal} {Light: Science \& Applications}\
  }\textbf {\bibinfo {volume} {8}},\ \bibinfo {pages} {40} (\bibinfo {year}
  {2019})}\BibitemShut {NoStop}%
\bibitem [{\citenamefont {Busch}\ \emph {et~al.}(2007)\citenamefont {Busch},
  \citenamefont {von Freymann}, \citenamefont {Linden}, \citenamefont
  {Mingaleev}, \citenamefont {Tkeshelashvili},\ and\ \citenamefont
  {Wegener}}]{Busch2007Periodic}%
  \BibitemOpen
  \bibfield  {author} {\bibinfo {author} {\bibfnamefont {K.}~\bibnamefont
  {Busch}}, \bibinfo {author} {\bibfnamefont {G.}~\bibnamefont {von Freymann}},
  \bibinfo {author} {\bibfnamefont {S.}~\bibnamefont {Linden}}, \bibinfo
  {author} {\bibfnamefont {S.~F.}\ \bibnamefont {Mingaleev}}, \bibinfo {author}
  {\bibfnamefont {L.}~\bibnamefont {Tkeshelashvili}},\ and\ \bibinfo {author}
  {\bibfnamefont {M.}~\bibnamefont {Wegener}},\ }\bibfield  {title} {\bibinfo
  {title} {Periodic nanostructures for photonics},\ }\href
  {https://doi.org/10.1016/j.physrep.2007.02.011} {\bibfield  {journal}
  {\bibinfo  {journal} {Phys. Rep.}\ }\textbf {\bibinfo {volume} {444}},\
  \bibinfo {pages} {101} (\bibinfo {year} {2007})}\BibitemShut {NoStop}%
\bibitem [{\citenamefont {Li}\ \emph {et~al.}(2026{\natexlab{a}})\citenamefont
  {Li}, \citenamefont {Li}, \citenamefont {Chen}, \citenamefont {Xiao},
  \citenamefont {Huang}, \citenamefont {Zhuo}, \citenamefont {Guan},
  \citenamefont {Gao}, \citenamefont {Zhan}, \citenamefont {Lu},\ and\
  \citenamefont {Xie}}]{Li2026TightBinding}%
  \BibitemOpen
  \bibfield  {author} {\bibinfo {author} {\bibfnamefont {J.}~\bibnamefont
  {Li}}, \bibinfo {author} {\bibfnamefont {A.}~\bibnamefont {Li}}, \bibinfo
  {author} {\bibfnamefont {Y.}~\bibnamefont {Chen}}, \bibinfo {author}
  {\bibfnamefont {T.}~\bibnamefont {Xiao}}, \bibinfo {author} {\bibfnamefont
  {R.}~\bibnamefont {Huang}}, \bibinfo {author} {\bibfnamefont
  {X.}~\bibnamefont {Zhuo}}, \bibinfo {author} {\bibfnamefont {J.}~\bibnamefont
  {Guan}}, \bibinfo {author} {\bibfnamefont {Z.}~\bibnamefont {Gao}}, \bibinfo
  {author} {\bibfnamefont {P.}~\bibnamefont {Zhan}}, \bibinfo {author}
  {\bibfnamefont {M.}~\bibnamefont {Lu}},\ and\ \bibinfo {author}
  {\bibfnamefont {B.}~\bibnamefont {Xie}},\ }\bibfield  {title} {\bibinfo
  {title} {Tight-binding photonics},\ }\href
  {https://doi.org/10.1117/1.ap.8.5.054002} {\bibfield  {journal} {\bibinfo
  {journal} {Adv. Photonics}\ }\textbf {\bibinfo {volume} {8}},\ \bibinfo
  {pages} {054002} (\bibinfo {year} {2026}{\natexlab{a}})}\BibitemShut
  {NoStop}%
\bibitem [{\citenamefont {Lidorikis}\ \emph {et~al.}(1998)\citenamefont
  {Lidorikis}, \citenamefont {Sigalas}, \citenamefont {Economou},\ and\
  \citenamefont {Soukoulis}}]{Lidorikis1998TightBinding}%
  \BibitemOpen
  \bibfield  {author} {\bibinfo {author} {\bibfnamefont {E.}~\bibnamefont
  {Lidorikis}}, \bibinfo {author} {\bibfnamefont {M.~M.}\ \bibnamefont
  {Sigalas}}, \bibinfo {author} {\bibfnamefont {E.~N.}\ \bibnamefont
  {Economou}},\ and\ \bibinfo {author} {\bibfnamefont {C.~M.}\ \bibnamefont
  {Soukoulis}},\ }\bibfield  {title} {\bibinfo {title} {Tight-binding
  parametrization for photonic band gap materials},\ }\href
  {https://doi.org/10.1103/PhysRevLett.81.1405} {\bibfield  {journal} {\bibinfo
   {journal} {Phys. Rev. Lett.}\ }\textbf {\bibinfo {volume} {81}},\ \bibinfo
  {pages} {1405} (\bibinfo {year} {1998})}\BibitemShut {NoStop}%
\bibitem [{\citenamefont {Ozawa}\ \emph {et~al.}(2019)\citenamefont {Ozawa},
  \citenamefont {Price}, \citenamefont {Amo}, \citenamefont {Goldman},
  \citenamefont {Hafezi}, \citenamefont {Lu}, \citenamefont {Rechtsman},
  \citenamefont {Schuster}, \citenamefont {Simon}, \citenamefont {Zilberberg},\
  and\ \citenamefont {Carusotto}}]{Ozawa2019Topological}%
  \BibitemOpen
  \bibfield  {author} {\bibinfo {author} {\bibfnamefont {T.}~\bibnamefont
  {Ozawa}}, \bibinfo {author} {\bibfnamefont {H.~M.}\ \bibnamefont {Price}},
  \bibinfo {author} {\bibfnamefont {A.}~\bibnamefont {Amo}}, \bibinfo {author}
  {\bibfnamefont {N.}~\bibnamefont {Goldman}}, \bibinfo {author} {\bibfnamefont
  {M.}~\bibnamefont {Hafezi}}, \bibinfo {author} {\bibfnamefont
  {L.}~\bibnamefont {Lu}}, \bibinfo {author} {\bibfnamefont {M.~C.}\
  \bibnamefont {Rechtsman}}, \bibinfo {author} {\bibfnamefont {D.}~\bibnamefont
  {Schuster}}, \bibinfo {author} {\bibfnamefont {J.}~\bibnamefont {Simon}},
  \bibinfo {author} {\bibfnamefont {O.}~\bibnamefont {Zilberberg}},\ and\
  \bibinfo {author} {\bibfnamefont {I.}~\bibnamefont {Carusotto}},\ }\bibfield
  {title} {\bibinfo {title} {Topological photonics},\ }\href
  {https://doi.org/10.1103/RevModPhys.91.015006} {\bibfield  {journal}
  {\bibinfo  {journal} {Rev. Mod. Phys.}\ }\textbf {\bibinfo {volume} {91}},\
  \bibinfo {pages} {015006} (\bibinfo {year} {2019})}\BibitemShut {NoStop}%
\bibitem [{\citenamefont {Zhang}\ \emph {et~al.}(2023)\citenamefont {Zhang},
  \citenamefont {Zangeneh-Nejad}, \citenamefont {Chen}, \citenamefont {Lu},\
  and\ \citenamefont {Christensen}}]{Zhang2023SecondWave}%
  \BibitemOpen
  \bibfield  {author} {\bibinfo {author} {\bibfnamefont {X.}~\bibnamefont
  {Zhang}}, \bibinfo {author} {\bibfnamefont {F.}~\bibnamefont
  {Zangeneh-Nejad}}, \bibinfo {author} {\bibfnamefont {Z.-G.}\ \bibnamefont
  {Chen}}, \bibinfo {author} {\bibfnamefont {M.-H.}\ \bibnamefont {Lu}},\ and\
  \bibinfo {author} {\bibfnamefont {J.}~\bibnamefont {Christensen}},\
  }\bibfield  {title} {\bibinfo {title} {A second wave of topological phenomena
  in photonics and acoustics},\ }\href
  {https://doi.org/10.1038/s41586-023-06163-9} {\bibfield  {journal} {\bibinfo
  {journal} {Nature}\ }\textbf {\bibinfo {volume} {618}},\ \bibinfo {pages}
  {687} (\bibinfo {year} {2023})}\BibitemShut {NoStop}%
\bibitem [{\citenamefont {Leykam}\ \emph {et~al.}(2026)\citenamefont {Leykam},
  \citenamefont {Xue}, \citenamefont {Zhang},\ and\ \citenamefont
  {Chong}}]{Leykam2026Limitations}%
  \BibitemOpen
  \bibfield  {author} {\bibinfo {author} {\bibfnamefont {D.}~\bibnamefont
  {Leykam}}, \bibinfo {author} {\bibfnamefont {H.}~\bibnamefont {Xue}},
  \bibinfo {author} {\bibfnamefont {B.}~\bibnamefont {Zhang}},\ and\ \bibinfo
  {author} {\bibfnamefont {Y.~D.}\ \bibnamefont {Chong}},\ }\bibfield  {title}
  {\bibinfo {title} {Limitations and possibilities of topological photonics},\
  }\href {https://doi.org/10.1038/s42254-025-00889-3} {\bibfield  {journal}
  {\bibinfo  {journal} {Nat. Rev. Phys.}\ }\textbf {\bibinfo {volume} {8}},\
  \bibinfo {pages} {55} (\bibinfo {year} {2026})}\BibitemShut {NoStop}%
\bibitem [{\citenamefont {Hasan}\ and\ \citenamefont
  {Kane}(2010)}]{Hasan2010Colloquium}%
  \BibitemOpen
  \bibfield  {author} {\bibinfo {author} {\bibfnamefont {M.~Z.}\ \bibnamefont
  {Hasan}}\ and\ \bibinfo {author} {\bibfnamefont {C.~L.}\ \bibnamefont
  {Kane}},\ }\bibfield  {title} {\bibinfo {title} {Colloquium: Topological
  insulators},\ }\href {https://doi.org/10.1103/RevModPhys.82.3045} {\bibfield
  {journal} {\bibinfo  {journal} {Rev. Mod. Phys.}\ }\textbf {\bibinfo {volume}
  {82}},\ \bibinfo {pages} {3045} (\bibinfo {year} {2010})}\BibitemShut
  {NoStop}%
\bibitem [{\citenamefont {Qi}\ and\ \citenamefont
  {Zhang}(2011)}]{Qi2011Topological}%
  \BibitemOpen
  \bibfield  {author} {\bibinfo {author} {\bibfnamefont {X.-L.}\ \bibnamefont
  {Qi}}\ and\ \bibinfo {author} {\bibfnamefont {S.-C.}\ \bibnamefont {Zhang}},\
  }\bibfield  {title} {\bibinfo {title} {Topological insulators and
  superconductors},\ }\href {https://doi.org/10.1103/RevModPhys.83.1057}
  {\bibfield  {journal} {\bibinfo  {journal} {Rev. Mod. Phys.}\ }\textbf
  {\bibinfo {volume} {83}},\ \bibinfo {pages} {1057} (\bibinfo {year}
  {2011})}\BibitemShut {NoStop}%
\bibitem [{\citenamefont {Su}\ \emph {et~al.}(1979)\citenamefont {Su},
  \citenamefont {Schrieffer},\ and\ \citenamefont {Heeger}}]{Su1979Solitons}%
  \BibitemOpen
  \bibfield  {author} {\bibinfo {author} {\bibfnamefont {W.~P.}\ \bibnamefont
  {Su}}, \bibinfo {author} {\bibfnamefont {J.~R.}\ \bibnamefont {Schrieffer}},\
  and\ \bibinfo {author} {\bibfnamefont {A.~J.}\ \bibnamefont {Heeger}},\
  }\bibfield  {title} {\bibinfo {title} {Solitons in polyacetylene},\ }\href
  {https://doi.org/10.1103/PhysRevLett.42.1698} {\bibfield  {journal} {\bibinfo
   {journal} {Phys. Rev. Lett.}\ }\textbf {\bibinfo {volume} {42}},\ \bibinfo
  {pages} {1698} (\bibinfo {year} {1979})}\BibitemShut {NoStop}%
\bibitem [{\citenamefont {Haldane}(1988)}]{Haldane1988Model}%
  \BibitemOpen
  \bibfield  {author} {\bibinfo {author} {\bibfnamefont {F.~D.~M.}\
  \bibnamefont {Haldane}},\ }\bibfield  {title} {\bibinfo {title} {Model for a
  quantum {Hall} effect without {Landau} levels: Condensed-matter realization
  of the ``parity anomaly''},\ }\href
  {https://doi.org/10.1103/PhysRevLett.61.2015} {\bibfield  {journal} {\bibinfo
   {journal} {Phys. Rev. Lett.}\ }\textbf {\bibinfo {volume} {61}},\ \bibinfo
  {pages} {2015} (\bibinfo {year} {1988})}\BibitemShut {NoStop}%
\bibitem [{\citenamefont {Kane}\ and\ \citenamefont
  {Mele}(2005)}]{KaneMele2005QuantumSpinHall}%
  \BibitemOpen
  \bibfield  {author} {\bibinfo {author} {\bibfnamefont {C.~L.}\ \bibnamefont
  {Kane}}\ and\ \bibinfo {author} {\bibfnamefont {E.~J.}\ \bibnamefont
  {Mele}},\ }\bibfield  {title} {\bibinfo {title} {Quantum spin {Hall} effect
  in graphene},\ }\href {https://doi.org/10.1103/PhysRevLett.95.226801}
  {\bibfield  {journal} {\bibinfo  {journal} {Phys. Rev. Lett.}\ }\textbf
  {\bibinfo {volume} {95}},\ \bibinfo {pages} {226801} (\bibinfo {year}
  {2005})}\BibitemShut {NoStop}%
\bibitem [{\citenamefont {Bernevig}\ \emph {et~al.}(2006)\citenamefont
  {Bernevig}, \citenamefont {Hughes},\ and\ \citenamefont
  {Zhang}}]{Bernevig2006QuantumSpinHall}%
  \BibitemOpen
  \bibfield  {author} {\bibinfo {author} {\bibfnamefont {B.~A.}\ \bibnamefont
  {Bernevig}}, \bibinfo {author} {\bibfnamefont {T.~L.}\ \bibnamefont
  {Hughes}},\ and\ \bibinfo {author} {\bibfnamefont {S.-C.}\ \bibnamefont
  {Zhang}},\ }\bibfield  {title} {\bibinfo {title} {Quantum spin {Hall} effect
  and topological phase transition in {HgTe} quantum wells},\ }\href
  {https://doi.org/10.1126/science.1133734} {\bibfield  {journal} {\bibinfo
  {journal} {Science}\ }\textbf {\bibinfo {volume} {314}},\ \bibinfo {pages}
  {1757} (\bibinfo {year} {2006})}\BibitemShut {NoStop}%
\bibitem [{\citenamefont {Benalcazar}\ \emph {et~al.}(2017)\citenamefont
  {Benalcazar}, \citenamefont {Bernevig},\ and\ \citenamefont
  {Hughes}}]{Benalcazar2017Quantized}%
  \BibitemOpen
  \bibfield  {author} {\bibinfo {author} {\bibfnamefont {W.~A.}\ \bibnamefont
  {Benalcazar}}, \bibinfo {author} {\bibfnamefont {B.~A.}\ \bibnamefont
  {Bernevig}},\ and\ \bibinfo {author} {\bibfnamefont {T.~L.}\ \bibnamefont
  {Hughes}},\ }\bibfield  {title} {\bibinfo {title} {Quantized electric
  multipole insulators},\ }\href {https://doi.org/10.1126/science.aah6442}
  {\bibfield  {journal} {\bibinfo  {journal} {Science}\ }\textbf {\bibinfo
  {volume} {357}},\ \bibinfo {pages} {61} (\bibinfo {year} {2017})}\BibitemShut
  {NoStop}%
\bibitem [{\citenamefont {Bradlyn}\ \emph {et~al.}(2017)\citenamefont
  {Bradlyn}, \citenamefont {Elcoro}, \citenamefont {Cano}, \citenamefont
  {Vergniory}, \citenamefont {Wang}, \citenamefont {Felser}, \citenamefont
  {Aroyo},\ and\ \citenamefont {Bernevig}}]{Bradlyn2017Topological}%
  \BibitemOpen
  \bibfield  {author} {\bibinfo {author} {\bibfnamefont {B.}~\bibnamefont
  {Bradlyn}}, \bibinfo {author} {\bibfnamefont {L.}~\bibnamefont {Elcoro}},
  \bibinfo {author} {\bibfnamefont {J.}~\bibnamefont {Cano}}, \bibinfo {author}
  {\bibfnamefont {M.~G.}\ \bibnamefont {Vergniory}}, \bibinfo {author}
  {\bibfnamefont {Z.}~\bibnamefont {Wang}}, \bibinfo {author} {\bibfnamefont
  {C.}~\bibnamefont {Felser}}, \bibinfo {author} {\bibfnamefont {M.~I.}\
  \bibnamefont {Aroyo}},\ and\ \bibinfo {author} {\bibfnamefont {B.~A.}\
  \bibnamefont {Bernevig}},\ }\bibfield  {title} {\bibinfo {title} {Topological
  quantum chemistry},\ }\href {https://doi.org/10.1038/nature23268} {\bibfield
  {journal} {\bibinfo  {journal} {Nature}\ }\textbf {\bibinfo {volume} {547}},\
  \bibinfo {pages} {298} (\bibinfo {year} {2017})},\ \bibinfo {note} {author
  Correction: \href{https://doi.org/10.1038/s41586-020-2305-x}{Nature
  \textbf{582}, E14 (2020)}}\BibitemShut {NoStop}%
\bibitem [{\citenamefont {Li}\ \emph {et~al.}(2024)\citenamefont {Li},
  \citenamefont {Wang}, \citenamefont {Jia}, \citenamefont {Zhan},
  \citenamefont {Lu}, \citenamefont {Wang}, \citenamefont {Chen},\ and\
  \citenamefont {Xie}}]{Li2024Disentangled}%
  \BibitemOpen
  \bibfield  {author} {\bibinfo {author} {\bibfnamefont {J.}~\bibnamefont
  {Li}}, \bibinfo {author} {\bibfnamefont {H.}~\bibnamefont {Wang}}, \bibinfo
  {author} {\bibfnamefont {S.}~\bibnamefont {Jia}}, \bibinfo {author}
  {\bibfnamefont {P.}~\bibnamefont {Zhan}}, \bibinfo {author} {\bibfnamefont
  {M.}~\bibnamefont {Lu}}, \bibinfo {author} {\bibfnamefont {Z.}~\bibnamefont
  {Wang}}, \bibinfo {author} {\bibfnamefont {Y.}~\bibnamefont {Chen}},\ and\
  \bibinfo {author} {\bibfnamefont {B.-Y.}\ \bibnamefont {Xie}},\ }\bibfield
  {title} {\bibinfo {title} {Disentangled higher-orbital bands and chiral
  symmetric topology in confined {Mie} resonance photonic crystals},\ }\href
  {https://doi.org/10.1002/lpor.202300543} {\bibfield  {journal} {\bibinfo
  {journal} {Laser Photonics Rev.}\ }\textbf {\bibinfo {volume} {18}},\
  \bibinfo {pages} {2300543} (\bibinfo {year} {2024})}\BibitemShut {NoStop}%
\bibitem [{\citenamefont {Schulz}\ \emph {et~al.}(2022)\citenamefont {Schulz},
  \citenamefont {Noh}, \citenamefont {Benalcazar}, \citenamefont {Bahl},\ and\
  \citenamefont {von Freymann}}]{Schulz2022PhotonicQuadrupole}%
  \BibitemOpen
  \bibfield  {author} {\bibinfo {author} {\bibfnamefont {J.}~\bibnamefont
  {Schulz}}, \bibinfo {author} {\bibfnamefont {J.}~\bibnamefont {Noh}},
  \bibinfo {author} {\bibfnamefont {W.~A.}\ \bibnamefont {Benalcazar}},
  \bibinfo {author} {\bibfnamefont {G.}~\bibnamefont {Bahl}},\ and\ \bibinfo
  {author} {\bibfnamefont {G.}~\bibnamefont {von Freymann}},\ }\bibfield
  {title} {\bibinfo {title} {Photonic quadrupole topological insulator using
  orbital-induced synthetic flux},\ }\href
  {https://doi.org/10.1038/s41467-022-33894-6} {\bibfield  {journal} {\bibinfo
  {journal} {Nat. Commun.}\ }\textbf {\bibinfo {volume} {13}},\ \bibinfo
  {pages} {6597} (\bibinfo {year} {2022})}\BibitemShut {NoStop}%
\bibitem [{\citenamefont {Rechtsman}\ \emph {et~al.}(2013)\citenamefont
  {Rechtsman}, \citenamefont {Zeuner}, \citenamefont {Plotnik}, \citenamefont
  {Lumer}, \citenamefont {Podolsky}, \citenamefont {Dreisow}, \citenamefont
  {Nolte}, \citenamefont {Segev},\ and\ \citenamefont
  {Szameit}}]{Rechtsman2013Photonic}%
  \BibitemOpen
  \bibfield  {author} {\bibinfo {author} {\bibfnamefont {M.~C.}\ \bibnamefont
  {Rechtsman}}, \bibinfo {author} {\bibfnamefont {J.~M.}\ \bibnamefont
  {Zeuner}}, \bibinfo {author} {\bibfnamefont {Y.}~\bibnamefont {Plotnik}},
  \bibinfo {author} {\bibfnamefont {Y.}~\bibnamefont {Lumer}}, \bibinfo
  {author} {\bibfnamefont {D.}~\bibnamefont {Podolsky}}, \bibinfo {author}
  {\bibfnamefont {F.}~\bibnamefont {Dreisow}}, \bibinfo {author} {\bibfnamefont
  {S.}~\bibnamefont {Nolte}}, \bibinfo {author} {\bibfnamefont
  {M.}~\bibnamefont {Segev}},\ and\ \bibinfo {author} {\bibfnamefont
  {A.}~\bibnamefont {Szameit}},\ }\bibfield  {title} {\bibinfo {title}
  {Photonic {Floquet} topological insulators},\ }\href
  {https://doi.org/10.1038/nature12066} {\bibfield  {journal} {\bibinfo
  {journal} {Nature}\ }\textbf {\bibinfo {volume} {496}},\ \bibinfo {pages}
  {196} (\bibinfo {year} {2013})}\BibitemShut {NoStop}%
\bibitem [{\citenamefont {Maczewsky}\ \emph {et~al.}(2020)\citenamefont
  {Maczewsky}, \citenamefont {Heinrich}, \citenamefont {Kremer}, \citenamefont
  {Ivanov}, \citenamefont {Ehrhardt}, \citenamefont {Martinez}, \citenamefont
  {Kartashov}, \citenamefont {Konotop}, \citenamefont {Torner}, \citenamefont
  {Bauer},\ and\ \citenamefont {Szameit}}]{Maczewsky2020Nonlinearity}%
  \BibitemOpen
  \bibfield  {author} {\bibinfo {author} {\bibfnamefont {L.~J.}\ \bibnamefont
  {Maczewsky}}, \bibinfo {author} {\bibfnamefont {M.}~\bibnamefont {Heinrich}},
  \bibinfo {author} {\bibfnamefont {M.}~\bibnamefont {Kremer}}, \bibinfo
  {author} {\bibfnamefont {S.~K.}\ \bibnamefont {Ivanov}}, \bibinfo {author}
  {\bibfnamefont {M.}~\bibnamefont {Ehrhardt}}, \bibinfo {author}
  {\bibfnamefont {F.}~\bibnamefont {Martinez}}, \bibinfo {author}
  {\bibfnamefont {Y.~V.}\ \bibnamefont {Kartashov}}, \bibinfo {author}
  {\bibfnamefont {V.~V.}\ \bibnamefont {Konotop}}, \bibinfo {author}
  {\bibfnamefont {L.}~\bibnamefont {Torner}}, \bibinfo {author} {\bibfnamefont
  {D.}~\bibnamefont {Bauer}},\ and\ \bibinfo {author} {\bibfnamefont
  {A.}~\bibnamefont {Szameit}},\ }\bibfield  {title} {\bibinfo {title}
  {Nonlinearity-induced photonic topological insulator},\ }\href
  {https://doi.org/10.1126/science.abd2033} {\bibfield  {journal} {\bibinfo
  {journal} {Science}\ }\textbf {\bibinfo {volume} {370}},\ \bibinfo {pages}
  {701} (\bibinfo {year} {2020})}\BibitemShut {NoStop}%
\bibitem [{\citenamefont {Yariv}\ \emph {et~al.}(1999)\citenamefont {Yariv},
  \citenamefont {Xu}, \citenamefont {Lee},\ and\ \citenamefont
  {Scherer}}]{Yariv1999}%
  \BibitemOpen
  \bibfield  {author} {\bibinfo {author} {\bibfnamefont {A.}~\bibnamefont
  {Yariv}}, \bibinfo {author} {\bibfnamefont {Y.}~\bibnamefont {Xu}}, \bibinfo
  {author} {\bibfnamefont {R.~K.}\ \bibnamefont {Lee}},\ and\ \bibinfo {author}
  {\bibfnamefont {A.}~\bibnamefont {Scherer}},\ }\bibfield  {title} {\bibinfo
  {title} {Coupled-resonator optical waveguide: A proposal and analysis},\
  }\href {https://doi.org/10.1364/OL.24.000711} {\bibfield  {journal} {\bibinfo
   {journal} {Opt. Lett.}\ }\textbf {\bibinfo {volume} {24}},\ \bibinfo {pages}
  {711} (\bibinfo {year} {1999})}\BibitemShut {NoStop}%
\bibitem [{\citenamefont {Bayindir}\ \emph {et~al.}(2000)\citenamefont
  {Bayindir}, \citenamefont {Temelkuran},\ and\ \citenamefont
  {Ozbay}}]{Bayindir2000TightBinding}%
  \BibitemOpen
  \bibfield  {author} {\bibinfo {author} {\bibfnamefont {M.}~\bibnamefont
  {Bayindir}}, \bibinfo {author} {\bibfnamefont {B.}~\bibnamefont
  {Temelkuran}},\ and\ \bibinfo {author} {\bibfnamefont {E.}~\bibnamefont
  {Ozbay}},\ }\bibfield  {title} {\bibinfo {title} {Tight-binding description
  of the coupled defect modes in three-dimensional photonic crystals},\ }\href
  {https://doi.org/10.1103/PhysRevLett.84.2140} {\bibfield  {journal} {\bibinfo
   {journal} {Phys. Rev. Lett.}\ }\textbf {\bibinfo {volume} {84}},\ \bibinfo
  {pages} {2140} (\bibinfo {year} {2000})}\BibitemShut {NoStop}%
\bibitem [{\citenamefont {Dossou}\ \emph {et~al.}(2009)\citenamefont {Dossou},
  \citenamefont {Poulton}, \citenamefont {Botten}, \citenamefont {Mahmoodian},
  \citenamefont {McPhedran},\ and\ \citenamefont {de~Sterke}}]{Dossou2009}%
  \BibitemOpen
  \bibfield  {author} {\bibinfo {author} {\bibfnamefont {K.~B.}\ \bibnamefont
  {Dossou}}, \bibinfo {author} {\bibfnamefont {C.~G.}\ \bibnamefont {Poulton}},
  \bibinfo {author} {\bibfnamefont {L.~C.}\ \bibnamefont {Botten}}, \bibinfo
  {author} {\bibfnamefont {S.}~\bibnamefont {Mahmoodian}}, \bibinfo {author}
  {\bibfnamefont {R.~C.}\ \bibnamefont {McPhedran}},\ and\ \bibinfo {author}
  {\bibfnamefont {C.~M.}\ \bibnamefont {de~Sterke}},\ }\bibfield  {title}
  {\bibinfo {title} {Modes of symmetric composite defects in two-dimensional
  photonic crystals},\ }\href {https://doi.org/10.1103/PhysRevA.80.013826}
  {\bibfield  {journal} {\bibinfo  {journal} {Phys. Rev. A}\ }\textbf {\bibinfo
  {volume} {80}},\ \bibinfo {pages} {013826} (\bibinfo {year}
  {2009})}\BibitemShut {NoStop}%
\bibitem [{\citenamefont {Wang}\ \emph {et~al.}(2025)\citenamefont {Wang},
  \citenamefont {Meng}, \citenamefont {Yan}, \citenamefont {Zhao},
  \citenamefont {Yang}, \citenamefont {Chen}, \citenamefont {Cheng},
  \citenamefont {Xiao}, \citenamefont {Shum}, \citenamefont {Liu},
  \citenamefont {Yang}, \citenamefont {Chen}, \citenamefont {Xi}, \citenamefont
  {Zhu}, \citenamefont {Xie},\ and\ \citenamefont
  {Gao}}]{Wang2025ThreeDimensional}%
  \BibitemOpen
  \bibfield  {author} {\bibinfo {author} {\bibfnamefont {Z.}~\bibnamefont
  {Wang}}, \bibinfo {author} {\bibfnamefont {Y.}~\bibnamefont {Meng}}, \bibinfo
  {author} {\bibfnamefont {B.}~\bibnamefont {Yan}}, \bibinfo {author}
  {\bibfnamefont {D.}~\bibnamefont {Zhao}}, \bibinfo {author} {\bibfnamefont
  {L.}~\bibnamefont {Yang}}, \bibinfo {author} {\bibfnamefont {J.}~\bibnamefont
  {Chen}}, \bibinfo {author} {\bibfnamefont {M.}~\bibnamefont {Cheng}},
  \bibinfo {author} {\bibfnamefont {T.}~\bibnamefont {Xiao}}, \bibinfo {author}
  {\bibfnamefont {P.~P.}\ \bibnamefont {Shum}}, \bibinfo {author}
  {\bibfnamefont {G.-G.}\ \bibnamefont {Liu}}, \bibinfo {author} {\bibfnamefont
  {Y.}~\bibnamefont {Yang}}, \bibinfo {author} {\bibfnamefont {H.}~\bibnamefont
  {Chen}}, \bibinfo {author} {\bibfnamefont {X.}~\bibnamefont {Xi}}, \bibinfo
  {author} {\bibfnamefont {Z.-X.}\ \bibnamefont {Zhu}}, \bibinfo {author}
  {\bibfnamefont {B.}~\bibnamefont {Xie}},\ and\ \bibinfo {author}
  {\bibfnamefont {Z.}~\bibnamefont {Gao}},\ }\bibfield  {title} {\bibinfo
  {title} {Realization of a three-dimensional photonic higher-order topological
  insulator},\ }\href {https://doi.org/10.1038/s41467-025-58051-7} {\bibfield
  {journal} {\bibinfo  {journal} {Nat. Commun.}\ }\textbf {\bibinfo {volume}
  {16}},\ \bibinfo {pages} {3122} (\bibinfo {year} {2025})}\BibitemShut
  {NoStop}%
\bibitem [{\citenamefont {Huang}\ \emph {et~al.}(2025)\citenamefont {Huang},
  \citenamefont {Li}, \citenamefont {Jia}, \citenamefont {Hu}, \citenamefont
  {Li}, \citenamefont {Li}, \citenamefont {Xie}, \citenamefont {Lu},
  \citenamefont {Zhan}, \citenamefont {Chen},\ and\ \citenamefont
  {Wang}}]{Huang2025SyntheticFlux}%
  \BibitemOpen
  \bibfield  {author} {\bibinfo {author} {\bibfnamefont {R.}~\bibnamefont
  {Huang}}, \bibinfo {author} {\bibfnamefont {H.}~\bibnamefont {Li}}, \bibinfo
  {author} {\bibfnamefont {S.}~\bibnamefont {Jia}}, \bibinfo {author}
  {\bibfnamefont {J.}~\bibnamefont {Hu}}, \bibinfo {author} {\bibfnamefont
  {S.}~\bibnamefont {Li}}, \bibinfo {author} {\bibfnamefont {J.}~\bibnamefont
  {Li}}, \bibinfo {author} {\bibfnamefont {B.}~\bibnamefont {Xie}}, \bibinfo
  {author} {\bibfnamefont {M.}~\bibnamefont {Lu}}, \bibinfo {author}
  {\bibfnamefont {P.}~\bibnamefont {Zhan}}, \bibinfo {author} {\bibfnamefont
  {Y.}~\bibnamefont {Chen}},\ and\ \bibinfo {author} {\bibfnamefont
  {Z.}~\bibnamefont {Wang}},\ }\bibfield  {title} {\bibinfo {title}
  {Experimental realization of synthetic $\pi$-flux photonic crystals},\ }\href
  {https://doi.org/10.1103/s4vq-hyck} {\bibfield  {journal} {\bibinfo
  {journal} {Phys. Rev. Lett.}\ }\textbf {\bibinfo {volume} {135}},\ \bibinfo
  {pages} {216603} (\bibinfo {year} {2025})}\BibitemShut {NoStop}%
\bibitem [{\citenamefont {Li}\ \emph {et~al.}(2026{\natexlab{b}})\citenamefont
  {Li}, \citenamefont {He}, \citenamefont {Wang}, \citenamefont {Jia},
  \citenamefont {Li}, \citenamefont {Huang}, \citenamefont {Huang},
  \citenamefont {Cai}, \citenamefont {Lu}, \citenamefont {Xie}, \citenamefont
  {Zhan},\ and\ \citenamefont {Wang}}]{Li2026GaugeDelocalization}%
  \BibitemOpen
  \bibfield  {author} {\bibinfo {author} {\bibfnamefont {S.}~\bibnamefont
  {Li}}, \bibinfo {author} {\bibfnamefont {Y.}~\bibnamefont {He}}, \bibinfo
  {author} {\bibfnamefont {Y.}~\bibnamefont {Wang}}, \bibinfo {author}
  {\bibfnamefont {S.}~\bibnamefont {Jia}}, \bibinfo {author} {\bibfnamefont
  {H.}~\bibnamefont {Li}}, \bibinfo {author} {\bibfnamefont {R.}~\bibnamefont
  {Huang}}, \bibinfo {author} {\bibfnamefont {H.}~\bibnamefont {Huang}},
  \bibinfo {author} {\bibfnamefont {H.}~\bibnamefont {Cai}}, \bibinfo {author}
  {\bibfnamefont {M.}~\bibnamefont {Lu}}, \bibinfo {author} {\bibfnamefont
  {B.}~\bibnamefont {Xie}}, \bibinfo {author} {\bibfnamefont {P.}~\bibnamefont
  {Zhan}},\ and\ \bibinfo {author} {\bibfnamefont {Z.}~\bibnamefont {Wang}},\
  }\bibfield  {title} {\bibinfo {title} {Gauge-tunable uniform delocalization
  of higher-order topological photonic modes},\ }\href
  {https://doi.org/10.1103/dkvp-41zt} {\bibfield  {journal} {\bibinfo
  {journal} {Phys. Rev. Lett.}\ }\textbf {\bibinfo {volume} {136}},\ \bibinfo
  {pages} {133801} (\bibinfo {year} {2026}{\natexlab{b}})}\BibitemShut
  {NoStop}%
\bibitem [{\citenamefont {Yan}\ \emph {et~al.}(2025)\citenamefont {Yan},
  \citenamefont {Qi}, \citenamefont {Wang}, \citenamefont {Meng}, \citenamefont
  {Yang}, \citenamefont {Zhu}, \citenamefont {Chen}, \citenamefont {Zhong},
  \citenamefont {Cheng}, \citenamefont {Xi},\ and\ \citenamefont
  {Gao}}]{Yan2025DiracVortex}%
  \BibitemOpen
  \bibfield  {author} {\bibinfo {author} {\bibfnamefont {B.}~\bibnamefont
  {Yan}}, \bibinfo {author} {\bibfnamefont {Y.}~\bibnamefont {Qi}}, \bibinfo
  {author} {\bibfnamefont {Z.}~\bibnamefont {Wang}}, \bibinfo {author}
  {\bibfnamefont {Y.}~\bibnamefont {Meng}}, \bibinfo {author} {\bibfnamefont
  {L.}~\bibnamefont {Yang}}, \bibinfo {author} {\bibfnamefont {Z.-X.}\
  \bibnamefont {Zhu}}, \bibinfo {author} {\bibfnamefont {J.-M.}\ \bibnamefont
  {Chen}}, \bibinfo {author} {\bibfnamefont {Y.}~\bibnamefont {Zhong}},
  \bibinfo {author} {\bibfnamefont {M.-Q.}\ \bibnamefont {Cheng}}, \bibinfo
  {author} {\bibfnamefont {X.}~\bibnamefont {Xi}},\ and\ \bibinfo {author}
  {\bibfnamefont {Z.}~\bibnamefont {Gao}},\ }\bibfield  {title} {\bibinfo
  {title} {Topological {Dirac}-vortex modes in a three-dimensional photonic
  topological insulator},\ }\href {https://doi.org/10.1038/s41467-025-61238-7}
  {\bibfield  {journal} {\bibinfo  {journal} {Nat. Commun.}\ }\textbf {\bibinfo
  {volume} {16}},\ \bibinfo {pages} {5659} (\bibinfo {year}
  {2025})}\BibitemShut {NoStop}%
\bibitem [{\citenamefont {Douglas}\ \emph {et~al.}(2015)\citenamefont
  {Douglas}, \citenamefont {Habibian}, \citenamefont {Hung}, \citenamefont
  {Gorshkov}, \citenamefont {Kimble},\ and\ \citenamefont
  {Chang}}]{Douglas2015QuantumManyBody}%
  \BibitemOpen
  \bibfield  {author} {\bibinfo {author} {\bibfnamefont {J.~S.}\ \bibnamefont
  {Douglas}}, \bibinfo {author} {\bibfnamefont {H.}~\bibnamefont {Habibian}},
  \bibinfo {author} {\bibfnamefont {C.-L.}\ \bibnamefont {Hung}}, \bibinfo
  {author} {\bibfnamefont {A.~V.}\ \bibnamefont {Gorshkov}}, \bibinfo {author}
  {\bibfnamefont {H.~J.}\ \bibnamefont {Kimble}},\ and\ \bibinfo {author}
  {\bibfnamefont {D.~E.}\ \bibnamefont {Chang}},\ }\bibfield  {title} {\bibinfo
  {title} {Quantum many-body models with cold atoms coupled to photonic
  crystals},\ }\href {https://doi.org/10.1038/nphoton.2015.57} {\bibfield
  {journal} {\bibinfo  {journal} {Nat. Photonics}\ }\textbf {\bibinfo {volume}
  {9}},\ \bibinfo {pages} {326} (\bibinfo {year} {2015})}\BibitemShut {NoStop}%
\bibitem [{\citenamefont {Hood}\ \emph {et~al.}(2016)\citenamefont {Hood},
  \citenamefont {Goban}, \citenamefont {Asenjo-Garcia}, \citenamefont {Lu},
  \citenamefont {Yu}, \citenamefont {Chang},\ and\ \citenamefont
  {Kimble}}]{Hood2016BandEdge}%
  \BibitemOpen
  \bibfield  {author} {\bibinfo {author} {\bibfnamefont {J.~D.}\ \bibnamefont
  {Hood}}, \bibinfo {author} {\bibfnamefont {A.}~\bibnamefont {Goban}},
  \bibinfo {author} {\bibfnamefont {A.}~\bibnamefont {Asenjo-Garcia}}, \bibinfo
  {author} {\bibfnamefont {M.}~\bibnamefont {Lu}}, \bibinfo {author}
  {\bibfnamefont {S.-P.}\ \bibnamefont {Yu}}, \bibinfo {author} {\bibfnamefont
  {D.~E.}\ \bibnamefont {Chang}},\ and\ \bibinfo {author} {\bibfnamefont
  {H.~J.}\ \bibnamefont {Kimble}},\ }\bibfield  {title} {\bibinfo {title}
  {Atom--atom interactions around the band edge of a photonic crystal
  waveguide},\ }\href {https://doi.org/10.1073/pnas.1603788113} {\bibfield
  {journal} {\bibinfo  {journal} {Proc. Natl. Acad. Sci. U.S.A.}\ }\textbf
  {\bibinfo {volume} {113}},\ \bibinfo {pages} {10507} (\bibinfo {year}
  {2016})}\BibitemShut {NoStop}%
\bibitem [{\citenamefont {Guo}\ \emph {et~al.}(2026)\citenamefont {Guo},
  \citenamefont {Huang}, \citenamefont {Tian}, \citenamefont {Zhou},
  \citenamefont {Wang},\ and\ \citenamefont {Zhou}}]{Guo2026DefectEvolved}%
  \BibitemOpen
  \bibfield  {author} {\bibinfo {author} {\bibfnamefont {S.}~\bibnamefont
  {Guo}}, \bibinfo {author} {\bibfnamefont {W.}~\bibnamefont {Huang}}, \bibinfo
  {author} {\bibfnamefont {F.}~\bibnamefont {Tian}}, \bibinfo {author}
  {\bibfnamefont {Y.}~\bibnamefont {Zhou}}, \bibinfo {author} {\bibfnamefont
  {Y.}~\bibnamefont {Wang}},\ and\ \bibinfo {author} {\bibfnamefont
  {T.}~\bibnamefont {Zhou}},\ }\bibfield  {title} {\bibinfo {title}
  {Defect-evolved quadrupole higher-order topological nanolasers},\ }\href
  {https://doi.org/10.1038/s41467-026-70056-4} {\bibfield  {journal} {\bibinfo
  {journal} {Nat. Commun.}\ }\textbf {\bibinfo {volume} {17}},\ \bibinfo
  {pages} {3238} (\bibinfo {year} {2026})}\BibitemShut {NoStop}%
\bibitem [{\citenamefont {Serra-Garcia}\ \emph {et~al.}(2018)\citenamefont
  {Serra-Garcia}, \citenamefont {Peri}, \citenamefont {S{\"u}sstrunk},
  \citenamefont {Bilal}, \citenamefont {Larsen}, \citenamefont {Villanueva},\
  and\ \citenamefont {Huber}}]{SerraGarcia2018Phononic}%
  \BibitemOpen
  \bibfield  {author} {\bibinfo {author} {\bibfnamefont {M.}~\bibnamefont
  {Serra-Garcia}}, \bibinfo {author} {\bibfnamefont {V.}~\bibnamefont {Peri}},
  \bibinfo {author} {\bibfnamefont {R.}~\bibnamefont {S{\"u}sstrunk}}, \bibinfo
  {author} {\bibfnamefont {O.~R.}\ \bibnamefont {Bilal}}, \bibinfo {author}
  {\bibfnamefont {T.}~\bibnamefont {Larsen}}, \bibinfo {author} {\bibfnamefont
  {L.~G.}\ \bibnamefont {Villanueva}},\ and\ \bibinfo {author} {\bibfnamefont
  {S.~D.}\ \bibnamefont {Huber}},\ }\bibfield  {title} {\bibinfo {title}
  {Observation of a phononic quadrupole topological insulator},\ }\href
  {https://doi.org/10.1038/nature25156} {\bibfield  {journal} {\bibinfo
  {journal} {Nature}\ }\textbf {\bibinfo {volume} {555}},\ \bibinfo {pages}
  {342} (\bibinfo {year} {2018})}\BibitemShut {NoStop}%
\bibitem [{\citenamefont {Peterson}\ \emph {et~al.}(2018)\citenamefont
  {Peterson}, \citenamefont {Benalcazar}, \citenamefont {Hughes},\ and\
  \citenamefont {Bahl}}]{Peterson2018MicrowaveQuadrupole}%
  \BibitemOpen
  \bibfield  {author} {\bibinfo {author} {\bibfnamefont {C.~W.}\ \bibnamefont
  {Peterson}}, \bibinfo {author} {\bibfnamefont {W.~A.}\ \bibnamefont
  {Benalcazar}}, \bibinfo {author} {\bibfnamefont {T.~L.}\ \bibnamefont
  {Hughes}},\ and\ \bibinfo {author} {\bibfnamefont {G.}~\bibnamefont {Bahl}},\
  }\bibfield  {title} {\bibinfo {title} {A quantized microwave quadrupole
  insulator with topologically protected corner states},\ }\href
  {https://doi.org/10.1038/nature25777} {\bibfield  {journal} {\bibinfo
  {journal} {Nature}\ }\textbf {\bibinfo {volume} {555}},\ \bibinfo {pages}
  {346} (\bibinfo {year} {2018})}\BibitemShut {NoStop}%
\bibitem [{\citenamefont {Imhof}\ \emph {et~al.}(2018)\citenamefont {Imhof},
  \citenamefont {Berger}, \citenamefont {Bayer}, \citenamefont {Brehm},
  \citenamefont {Molenkamp}, \citenamefont {Kiessling}, \citenamefont
  {Schindler}, \citenamefont {Lee}, \citenamefont {Greiter}, \citenamefont
  {Neupert},\ and\ \citenamefont {Thomale}}]{Imhof2018Topolectrical}%
  \BibitemOpen
  \bibfield  {author} {\bibinfo {author} {\bibfnamefont {S.}~\bibnamefont
  {Imhof}}, \bibinfo {author} {\bibfnamefont {C.}~\bibnamefont {Berger}},
  \bibinfo {author} {\bibfnamefont {F.}~\bibnamefont {Bayer}}, \bibinfo
  {author} {\bibfnamefont {J.}~\bibnamefont {Brehm}}, \bibinfo {author}
  {\bibfnamefont {L.~W.}\ \bibnamefont {Molenkamp}}, \bibinfo {author}
  {\bibfnamefont {T.}~\bibnamefont {Kiessling}}, \bibinfo {author}
  {\bibfnamefont {F.}~\bibnamefont {Schindler}}, \bibinfo {author}
  {\bibfnamefont {C.~H.}\ \bibnamefont {Lee}}, \bibinfo {author} {\bibfnamefont
  {M.}~\bibnamefont {Greiter}}, \bibinfo {author} {\bibfnamefont
  {T.}~\bibnamefont {Neupert}},\ and\ \bibinfo {author} {\bibfnamefont
  {R.}~\bibnamefont {Thomale}},\ }\bibfield  {title} {\bibinfo {title}
  {Topolectrical-circuit realization of topological corner modes},\ }\href
  {https://doi.org/10.1038/s41567-018-0246-1} {\bibfield  {journal} {\bibinfo
  {journal} {Nat. Phys.}\ }\textbf {\bibinfo {volume} {14}},\ \bibinfo {pages}
  {925} (\bibinfo {year} {2018})}\BibitemShut {NoStop}%
\bibitem [{\citenamefont {Mittal}\ \emph {et~al.}(2019)\citenamefont {Mittal},
  \citenamefont {Orre}, \citenamefont {Zhu}, \citenamefont {Gorlach},
  \citenamefont {Poddubny},\ and\ \citenamefont
  {Hafezi}}]{Mittal2019Quadrupole}%
  \BibitemOpen
  \bibfield  {author} {\bibinfo {author} {\bibfnamefont {S.}~\bibnamefont
  {Mittal}}, \bibinfo {author} {\bibfnamefont {V.~V.}\ \bibnamefont {Orre}},
  \bibinfo {author} {\bibfnamefont {G.}~\bibnamefont {Zhu}}, \bibinfo {author}
  {\bibfnamefont {M.~A.}\ \bibnamefont {Gorlach}}, \bibinfo {author}
  {\bibfnamefont {A.}~\bibnamefont {Poddubny}},\ and\ \bibinfo {author}
  {\bibfnamefont {M.}~\bibnamefont {Hafezi}},\ }\bibfield  {title} {\bibinfo
  {title} {Photonic quadrupole topological phases},\ }\href
  {https://doi.org/10.1038/s41566-019-0452-0} {\bibfield  {journal} {\bibinfo
  {journal} {Nat. Photonics}\ }\textbf {\bibinfo {volume} {13}},\ \bibinfo
  {pages} {692} (\bibinfo {year} {2019})}\BibitemShut {NoStop}%
\bibitem [{\citenamefont {Kee}\ \emph {et~al.}(2003)\citenamefont {Kee},
  \citenamefont {Lim},\ and\ \citenamefont {Lee}}]{Kee2003}%
  \BibitemOpen
  \bibfield  {author} {\bibinfo {author} {\bibfnamefont {C.-S.}\ \bibnamefont
  {Kee}}, \bibinfo {author} {\bibfnamefont {H.}~\bibnamefont {Lim}},\ and\
  \bibinfo {author} {\bibfnamefont {J.}~\bibnamefont {Lee}},\ }\bibfield
  {title} {\bibinfo {title} {Coupling characteristics of localized photons in
  two-dimensional photonic crystals},\ }\href
  {https://doi.org/10.1103/PhysRevB.67.073103} {\bibfield  {journal} {\bibinfo
  {journal} {Phys. Rev. B}\ }\textbf {\bibinfo {volume} {67}},\ \bibinfo
  {pages} {073103} (\bibinfo {year} {2003})}\BibitemShut {NoStop}%
\bibitem [{\citenamefont {Haddadi}\ \emph {et~al.}(2014)\citenamefont
  {Haddadi}, \citenamefont {Hamel}, \citenamefont {Beaudoin}, \citenamefont
  {Sagnes}, \citenamefont {Sauvan}, \citenamefont {Lalanne}, \citenamefont
  {Levenson},\ and\ \citenamefont {Yacomotti}}]{Haddadi2014}%
  \BibitemOpen
  \bibfield  {author} {\bibinfo {author} {\bibfnamefont {S.}~\bibnamefont
  {Haddadi}}, \bibinfo {author} {\bibfnamefont {P.}~\bibnamefont {Hamel}},
  \bibinfo {author} {\bibfnamefont {G.}~\bibnamefont {Beaudoin}}, \bibinfo
  {author} {\bibfnamefont {I.}~\bibnamefont {Sagnes}}, \bibinfo {author}
  {\bibfnamefont {C.}~\bibnamefont {Sauvan}}, \bibinfo {author} {\bibfnamefont
  {P.}~\bibnamefont {Lalanne}}, \bibinfo {author} {\bibfnamefont {J.~A.}\
  \bibnamefont {Levenson}},\ and\ \bibinfo {author} {\bibfnamefont {A.~M.}\
  \bibnamefont {Yacomotti}},\ }\bibfield  {title} {\bibinfo {title} {Photonic
  molecules: Tailoring the coupling strength and sign},\ }\href
  {https://doi.org/10.1364/OE.22.012359} {\bibfield  {journal} {\bibinfo
  {journal} {Opt. Express}\ }\textbf {\bibinfo {volume} {22}},\ \bibinfo
  {pages} {12359} (\bibinfo {year} {2014})}\BibitemShut {NoStop}%
\bibitem [{\citenamefont {Zheng}\ \emph {et~al.}(2026)\citenamefont {Zheng},
  \citenamefont {Ma}, \citenamefont {Tian},\ and\ \citenamefont
  {Yi}}]{Zheng2026Tailored}%
  \BibitemOpen
  \bibfield  {author} {\bibinfo {author} {\bibfnamefont {J.}~\bibnamefont
  {Zheng}}, \bibinfo {author} {\bibfnamefont {J.}~\bibnamefont {Ma}}, \bibinfo
  {author} {\bibfnamefont {Z.~Q.}\ \bibnamefont {Tian}},\ and\ \bibinfo
  {author} {\bibfnamefont {J.}~\bibnamefont {Yi}},\ }\bibfield  {title}
  {\bibinfo {title} {Tailored tight-binding couplings in photonic crystals},\
  }\href {https://doi.org/10.1002/lpor.202502889} {\bibfield  {journal}
  {\bibinfo  {journal} {Laser Photonics Rev.}\ }\textbf {\bibinfo {volume}
  {20}},\ \bibinfo {pages} {e02889} (\bibinfo {year} {2026})}\BibitemShut
  {NoStop}%
\bibitem [{\citenamefont {Johnson}\ and\ \citenamefont
  {Joannopoulos}(2000)}]{Johnson2000Layered}%
  \BibitemOpen
  \bibfield  {author} {\bibinfo {author} {\bibfnamefont {S.~G.}\ \bibnamefont
  {Johnson}}\ and\ \bibinfo {author} {\bibfnamefont {J.~D.}\ \bibnamefont
  {Joannopoulos}},\ }\bibfield  {title} {\bibinfo {title} {Three-dimensionally
  periodic dielectric layered structure with omnidirectional photonic band
  gap},\ }\href {https://doi.org/10.1063/1.1328369} {\bibfield  {journal}
  {\bibinfo  {journal} {Appl. Phys. Lett.}\ }\textbf {\bibinfo {volume} {77}},\
  \bibinfo {pages} {3490} (\bibinfo {year} {2000})}\BibitemShut {NoStop}%
\bibitem [{\citenamefont {Povinelli}\ \emph {et~al.}(2001)\citenamefont
  {Povinelli}, \citenamefont {Johnson}, \citenamefont {Fan},\ and\
  \citenamefont {Joannopoulos}}]{Povinelli2001Emulation}%
  \BibitemOpen
  \bibfield  {author} {\bibinfo {author} {\bibfnamefont {M.~L.}\ \bibnamefont
  {Povinelli}}, \bibinfo {author} {\bibfnamefont {S.~G.}\ \bibnamefont
  {Johnson}}, \bibinfo {author} {\bibfnamefont {S.}~\bibnamefont {Fan}},\ and\
  \bibinfo {author} {\bibfnamefont {J.~D.}\ \bibnamefont {Joannopoulos}},\
  }\bibfield  {title} {\bibinfo {title} {Emulation of two-dimensional photonic
  crystal defect modes in a photonic crystal with a three-dimensional photonic
  band gap},\ }\href {https://doi.org/10.1103/PhysRevB.64.075313} {\bibfield
  {journal} {\bibinfo  {journal} {Phys. Rev. B}\ }\textbf {\bibinfo {volume}
  {64}},\ \bibinfo {pages} {075313} (\bibinfo {year} {2001})}\BibitemShut
  {NoStop}%
\bibitem [{\citenamefont {Ezawa}(2018)}]{Ezawa2018Breathing}%
  \BibitemOpen
  \bibfield  {author} {\bibinfo {author} {\bibfnamefont {M.}~\bibnamefont
  {Ezawa}},\ }\bibfield  {title} {\bibinfo {title} {Higher-order topological
  insulators and semimetals on the breathing kagome and pyrochlore lattices},\
  }\href {https://doi.org/10.1103/PhysRevLett.120.026801} {\bibfield  {journal}
  {\bibinfo  {journal} {Phys. Rev. Lett.}\ }\textbf {\bibinfo {volume} {120}},\
  \bibinfo {pages} {026801} (\bibinfo {year} {2018})}\BibitemShut {NoStop}%
\bibitem [{\citenamefont {Ezawa}(2016)}]{Ezawa2016LoopNodal}%
  \BibitemOpen
  \bibfield  {author} {\bibinfo {author} {\bibfnamefont {M.}~\bibnamefont
  {Ezawa}},\ }\bibfield  {title} {\bibinfo {title} {Loop-nodal and point-nodal
  semimetals in three-dimensional honeycomb lattices},\ }\href
  {https://doi.org/10.1103/PhysRevLett.116.127202} {\bibfield  {journal}
  {\bibinfo  {journal} {Phys. Rev. Lett.}\ }\textbf {\bibinfo {volume} {116}},\
  \bibinfo {pages} {127202} (\bibinfo {year} {2016})}\BibitemShut {NoStop}%
\bibitem [{\citenamefont {Dossou}\ \emph {et~al.}(2008)\citenamefont {Dossou},
  \citenamefont {Botten}, \citenamefont {McPhedran}, \citenamefont {Poulton},
  \citenamefont {Asatryan},\ and\ \citenamefont
  {de~Sterke}}]{Dossou2008Shallow}%
  \BibitemOpen
  \bibfield  {author} {\bibinfo {author} {\bibfnamefont {K.~B.}\ \bibnamefont
  {Dossou}}, \bibinfo {author} {\bibfnamefont {L.~C.}\ \bibnamefont {Botten}},
  \bibinfo {author} {\bibfnamefont {R.~C.}\ \bibnamefont {McPhedran}}, \bibinfo
  {author} {\bibfnamefont {C.~G.}\ \bibnamefont {Poulton}}, \bibinfo {author}
  {\bibfnamefont {A.~A.}\ \bibnamefont {Asatryan}},\ and\ \bibinfo {author}
  {\bibfnamefont {C.~M.}\ \bibnamefont {de~Sterke}},\ }\bibfield  {title}
  {\bibinfo {title} {Shallow defect states in two-dimensional photonic
  crystals},\ }\href {https://doi.org/10.1103/PhysRevA.77.063839} {\bibfield
  {journal} {\bibinfo  {journal} {Phys. Rev. A}\ }\textbf {\bibinfo {volume}
  {77}},\ \bibinfo {pages} {063839} (\bibinfo {year} {2008})}\BibitemShut
  {NoStop}%
\bibitem [{\citenamefont {Silveirinha}(2020)}]{Silveirinha2021}%
  \BibitemOpen
  \bibfield  {author} {\bibinfo {author} {\bibfnamefont {M.~G.}\ \bibnamefont
  {Silveirinha}},\ }\bibfield  {title} {\bibinfo {title} {Modal expansions in
  dispersive material systems with application to quantum optics and
  topological photonics},\ }in\ \href {https://doi.org/10.1049/SBEW528E_ch24}
  {\emph {\bibinfo {booktitle} {Advances in Mathematical Methods for
  Electromagnetics}}},\ \bibinfo {editor} {edited by\ \bibinfo {editor}
  {\bibfnamefont {K.}~\bibnamefont {Kobayashi}}\ and\ \bibinfo {editor}
  {\bibfnamefont {P.~D.}\ \bibnamefont {Smith}}}\ (\bibinfo  {publisher}
  {Institution of Engineering and Technology},\ \bibinfo {year} {2020})\ pp.\
  \bibinfo {pages} {599--620}\BibitemShut {NoStop}%
\bibitem [{Sup()}]{SupplementalMaterial}%
  \BibitemOpen
  \href@noop {} {}\bibinfo {note} {See the accompanying Supplemental Material
  for Bloch--Green theory, coupling extraction and
  tuning, experimental methods, frequency-resolved bulk and corner
  measurements, finite-height and open-surface calculations, topological
  invariants, and three-dimensional lattice parameters and spectra.}\BibitemShut {Stop}%
\bibitem [{\citenamefont {Waks}\ and\ \citenamefont
  {Vuckovic}(2005)}]{Waks2005CoupledMode}%
  \BibitemOpen
  \bibfield  {author} {\bibinfo {author} {\bibfnamefont {E.}~\bibnamefont
  {Waks}}\ and\ \bibinfo {author} {\bibfnamefont {J.}~\bibnamefont
  {Vuckovic}},\ }\bibfield  {title} {\bibinfo {title} {Coupled mode theory for
  photonic crystal cavity-waveguide interaction},\ }\href
  {https://doi.org/10.1364/OPEX.13.005064} {\bibfield  {journal} {\bibinfo
  {journal} {Opt. Express}\ }\textbf {\bibinfo {volume} {13}},\ \bibinfo
  {pages} {5064} (\bibinfo {year} {2005})}\BibitemShut {NoStop}%
\bibitem [{\citenamefont {Li}\ \emph {et~al.}(2018)\citenamefont {Li},
  \citenamefont {Umer},\ and\ \citenamefont
  {Gong}}]{Li2018CornerConfigurations}%
  \BibitemOpen
  \bibfield  {author} {\bibinfo {author} {\bibfnamefont {L.}~\bibnamefont
  {Li}}, \bibinfo {author} {\bibfnamefont {M.}~\bibnamefont {Umer}},\ and\
  \bibinfo {author} {\bibfnamefont {J.}~\bibnamefont {Gong}},\ }\bibfield
  {title} {\bibinfo {title} {Direct prediction of corner state configurations
  from edge winding numbers in two- and three-dimensional chiral-symmetric
  lattice systems},\ }\href {https://doi.org/10.1103/PhysRevB.98.205422}
  {\bibfield  {journal} {\bibinfo  {journal} {Phys. Rev. B}\ }\textbf {\bibinfo
  {volume} {98}},\ \bibinfo {pages} {205422} (\bibinfo {year}
  {2018})}\BibitemShut {NoStop}%
\bibitem [{\citenamefont {Benalcazar}\ and\ \citenamefont
  {Cerjan}(2022)}]{Benalcazar2022Chiral}%
  \BibitemOpen
  \bibfield  {author} {\bibinfo {author} {\bibfnamefont {W.~A.}\ \bibnamefont
  {Benalcazar}}\ and\ \bibinfo {author} {\bibfnamefont {A.}~\bibnamefont
  {Cerjan}},\ }\bibfield  {title} {\bibinfo {title} {Chiral-symmetric
  higher-order topological phases of matter},\ }\href
  {https://doi.org/10.1103/PhysRevLett.128.127601} {\bibfield  {journal}
  {\bibinfo  {journal} {Phys. Rev. Lett.}\ }\textbf {\bibinfo {volume} {128}},\
  \bibinfo {pages} {127601} (\bibinfo {year} {2022})}\BibitemShut {NoStop}%
\bibitem [{\citenamefont {Corticelli}\ \emph {et~al.}(2022)\citenamefont
  {Corticelli}, \citenamefont {Moessner},\ and\ \citenamefont
  {McClarty}}]{Corticelli2022SpinSpace}%
  \BibitemOpen
  \bibfield  {author} {\bibinfo {author} {\bibfnamefont {A.}~\bibnamefont
  {Corticelli}}, \bibinfo {author} {\bibfnamefont {R.}~\bibnamefont
  {Moessner}},\ and\ \bibinfo {author} {\bibfnamefont {P.~A.}\ \bibnamefont
  {McClarty}},\ }\bibfield  {title} {\bibinfo {title} {Spin-space groups and
  magnon band topology},\ }\href {https://doi.org/10.1103/PhysRevB.105.064430}
  {\bibfield  {journal} {\bibinfo  {journal} {Phys. Rev. B}\ }\textbf {\bibinfo
  {volume} {105}},\ \bibinfo {pages} {064430} (\bibinfo {year}
  {2022})}\BibitemShut {NoStop}%
\bibitem [{\citenamefont {Wawrzik}\ \emph {et~al.}(2018)\citenamefont
  {Wawrzik}, \citenamefont {Lindner}, \citenamefont {Hermanns},\ and\
  \citenamefont {Trebst}}]{Wawrzik2018Honeycomb}%
  \BibitemOpen
  \bibfield  {author} {\bibinfo {author} {\bibfnamefont {D.}~\bibnamefont
  {Wawrzik}}, \bibinfo {author} {\bibfnamefont {D.}~\bibnamefont {Lindner}},
  \bibinfo {author} {\bibfnamefont {M.}~\bibnamefont {Hermanns}},\ and\
  \bibinfo {author} {\bibfnamefont {S.}~\bibnamefont {Trebst}},\ }\bibfield
  {title} {\bibinfo {title} {Topological semimetals and insulators in
  three-dimensional honeycomb materials},\ }\href
  {https://doi.org/10.1103/PhysRevB.98.115114} {\bibfield  {journal} {\bibinfo
  {journal} {Phys. Rev. B}\ }\textbf {\bibinfo {volume} {98}},\ \bibinfo
  {pages} {115114} (\bibinfo {year} {2018})}\BibitemShut {NoStop}%
\bibitem [{\citenamefont {Lemoult}\ \emph {et~al.}(2013)\citenamefont
  {Lemoult}, \citenamefont {Kaina}, \citenamefont {Fink},\ and\ \citenamefont
  {Lerosey}}]{Lemoult2013WavePropagation}%
  \BibitemOpen
  \bibfield  {author} {\bibinfo {author} {\bibfnamefont {F.}~\bibnamefont
  {Lemoult}}, \bibinfo {author} {\bibfnamefont {N.}~\bibnamefont {Kaina}},
  \bibinfo {author} {\bibfnamefont {M.}~\bibnamefont {Fink}},\ and\ \bibinfo
  {author} {\bibfnamefont {G.}~\bibnamefont {Lerosey}},\ }\bibfield  {title}
  {\bibinfo {title} {Wave propagation control at the deep subwavelength scale
  in metamaterials},\ }\href {https://doi.org/10.1038/nphys2480} {\bibfield
  {journal} {\bibinfo  {journal} {Nat. Phys.}\ }\textbf {\bibinfo {volume}
  {9}},\ \bibinfo {pages} {55} (\bibinfo {year} {2013})}\BibitemShut {NoStop}%
\bibitem [{\citenamefont {Miri}\ and\ \citenamefont
  {Al{\`u}}(2019)}]{Miri2019ExceptionalPoints}%
  \BibitemOpen
  \bibfield  {author} {\bibinfo {author} {\bibfnamefont {M.-A.}\ \bibnamefont
  {Miri}}\ and\ \bibinfo {author} {\bibfnamefont {A.}~\bibnamefont {Al{\`u}}},\
  }\bibfield  {title} {\bibinfo {title} {Exceptional points in optics and
  photonics},\ }\href {https://doi.org/10.1126/science.aar7709} {\bibfield
  {journal} {\bibinfo  {journal} {Science}\ }\textbf {\bibinfo {volume}
  {363}},\ \bibinfo {pages} {eaar7709} (\bibinfo {year} {2019})}\BibitemShut
  {NoStop}%
\bibitem [{\citenamefont {Yang}\ \emph {et~al.}(2023)\citenamefont {Yang},
  \citenamefont {Qin}, \citenamefont {Long}, \citenamefont {Yan}, \citenamefont
  {Yang}, \citenamefont {Li}, \citenamefont {Li}, \citenamefont {Hu},
  \citenamefont {Deng}, \citenamefont {Du},\ and\ \citenamefont
  {Bi}}]{Yang2023SelfBiased}%
  \BibitemOpen
  \bibfield  {author} {\bibinfo {author} {\bibfnamefont {W.}~\bibnamefont
  {Yang}}, \bibinfo {author} {\bibfnamefont {J.}~\bibnamefont {Qin}}, \bibinfo
  {author} {\bibfnamefont {J.}~\bibnamefont {Long}}, \bibinfo {author}
  {\bibfnamefont {W.}~\bibnamefont {Yan}}, \bibinfo {author} {\bibfnamefont
  {Y.}~\bibnamefont {Yang}}, \bibinfo {author} {\bibfnamefont {C.}~\bibnamefont
  {Li}}, \bibinfo {author} {\bibfnamefont {E.}~\bibnamefont {Li}}, \bibinfo
  {author} {\bibfnamefont {J.}~\bibnamefont {Hu}}, \bibinfo {author}
  {\bibfnamefont {L.}~\bibnamefont {Deng}}, \bibinfo {author} {\bibfnamefont
  {Q.}~\bibnamefont {Du}},\ and\ \bibinfo {author} {\bibfnamefont
  {L.}~\bibnamefont {Bi}},\ }\bibfield  {title} {\bibinfo {title} {A
  self-biased non-reciprocal magnetic metasurface for bidirectional phase
  modulation},\ }\href {https://doi.org/10.1038/s41928-023-00936-w} {\bibfield
  {journal} {\bibinfo  {journal} {Nat. Electron.}\ }\textbf {\bibinfo {volume}
  {6}},\ \bibinfo {pages} {225} (\bibinfo {year} {2023})}\BibitemShut {NoStop}%
\bibitem [{\citenamefont {Wang}\ \emph {et~al.}(2009)\citenamefont {Wang},
  \citenamefont {Chong}, \citenamefont {Joannopoulos},\ and\ \citenamefont
  {Solja{\v c}i{\'c}}}]{Wang2009Unidirectional}%
  \BibitemOpen
  \bibfield  {author} {\bibinfo {author} {\bibfnamefont {Z.}~\bibnamefont
  {Wang}}, \bibinfo {author} {\bibfnamefont {Y.}~\bibnamefont {Chong}},
  \bibinfo {author} {\bibfnamefont {J.~D.}\ \bibnamefont {Joannopoulos}},\ and\
  \bibinfo {author} {\bibfnamefont {M.}~\bibnamefont {Solja{\v c}i{\'c}}},\
  }\bibfield  {title} {\bibinfo {title} {Observation of unidirectional
  backscattering-immune topological electromagnetic states},\ }\href
  {https://doi.org/10.1038/nature08293} {\bibfield  {journal} {\bibinfo
  {journal} {Nature}\ }\textbf {\bibinfo {volume} {461}},\ \bibinfo {pages}
  {772} (\bibinfo {year} {2009})}\BibitemShut {NoStop}%
\bibitem [{\citenamefont {Hofstadter}(1976)}]{Hofstadter1976EnergyLevels}%
  \BibitemOpen
  \bibfield  {author} {\bibinfo {author} {\bibfnamefont {D.~R.}\ \bibnamefont
  {Hofstadter}},\ }\bibfield  {title} {\bibinfo {title} {Energy levels and wave
  functions of {Bloch} electrons in rational and irrational magnetic fields},\
  }\href {https://doi.org/10.1103/PhysRevB.14.2239} {\bibfield  {journal}
  {\bibinfo  {journal} {Phys. Rev. B}\ }\textbf {\bibinfo {volume} {14}},\
  \bibinfo {pages} {2239} (\bibinfo {year} {1976})}\BibitemShut {NoStop}%
\bibitem [{\citenamefont {John}\ and\ \citenamefont
  {Wang}(1990)}]{John1990QuantumElectrodynamics}%
  \BibitemOpen
  \bibfield  {author} {\bibinfo {author} {\bibfnamefont {S.}~\bibnamefont
  {John}}\ and\ \bibinfo {author} {\bibfnamefont {J.}~\bibnamefont {Wang}},\
  }\bibfield  {title} {\bibinfo {title} {Quantum electrodynamics near a
  photonic band gap: Photon bound states and dressed atoms},\ }\href
  {https://doi.org/10.1103/PhysRevLett.64.2418} {\bibfield  {journal} {\bibinfo
   {journal} {Phys. Rev. Lett.}\ }\textbf {\bibinfo {volume} {64}},\ \bibinfo
  {pages} {2418} (\bibinfo {year} {1990})}\BibitemShut {NoStop}%
\end{thebibliography}%

\clearpage
\section*{End Matter}

\textit{BBH Hamiltonian.---}
In the four-site basis of Fig.~\ref{fig:fig2}(d), the BBH Hamiltonian
\cite{Benalcazar2017Quantized} is
\begin{equation}
    H_{\mathrm{BBH}}( k) =
    \begin{pmatrix}
    0 & \tau_x & \tau_y & 0 \\
    \tau_x^* & 0 & 0 & -\tau_y \\
    \tau_y^* & 0 & 0 & \tau_x \\
    0 & -\tau_y^* & \tau_x^* & 0
    \end{pmatrix},
    \label{eq:bbh}
\end{equation}
where $\tau_x=t_{\mathrm{intra}}+t_{\mathrm{inter}}e^{-ik_x}$ and
$\tau_y=t_{\mathrm{intra}}+t_{\mathrm{inter}}e^{-ik_y}$.
Here $k_x,k_y$ are reduced Bloch coordinates equal to the phase advances
under the corresponding target-cell translations, and
$t_{\mathrm{intra}},t_{\mathrm{inter}}$ are intracell and intercell coupling
frequencies.  Eigenvalues are detunings from the common isolated resonance
frequency, and the minus signs encode $\pi$ flux.

\textit{Pyrochlore Hamiltonian.---}
The pyrochlore lattice \cite{Ezawa2018Breathing} has four couplings in the
layered BgV: $t_{\mathrm{intra}1},t_{\mathrm{intra}2}$ within cells and
$t_{\mathrm{inter}1},t_{\mathrm{inter}2}$ between cells, with indices
$1,2$ denoting lateral and oblique bonds [Fig.~\ref{fig:fig4}(c)].
The equal-separation realization has
$t_{\mathrm{intra}1}=t_{\mathrm{inter}1}=\SI{0.107}{GHz}$ and
$t_{\mathrm{intra}2}=t_{\mathrm{inter}2}=\SI{-0.012}{GHz}$.
Its Bloch Hamiltonian is
\begin{equation}
    H_{\mathrm{pyr}}( k) =
    \begin{pmatrix}
    0 & \tau_{12} & \tau_{13} & \tau_{14} \\
    \tau_{12}^* & 0 & \tau_{23} & \tau_{24} \\
    \tau_{13}^* & \tau_{23}^* & 0 & \tau_{34} \\
    \tau_{14}^* & \tau_{24}^* & \tau_{34}^* & 0
    \end{pmatrix}.
    \label{eq:pyrochlore}
\end{equation}
Here $\tau_{ij}=t_{\mathrm{intra},b_{ij}}+
t_{\mathrm{inter},b_{ij}}\exp[-i\beta_{ij}( k)]$, with
$b_{ij}=1$ for lateral bonds ($ij=23,24,34$) and
$b_{ij}=2$ for oblique bonds ($ij=12,13,14$) as
labeled in Fig.~\ref{fig:fig4}(c).  The phase factors follow the pyrochlore
Bloch convention:
$\beta_{12}=(k_x+k_y)/2$, $\beta_{13}=(k_y+k_z)/2$,
$\beta_{14}=(k_z+k_x)/2$, $\beta_{23}=(k_z-k_x)/2$,
$\beta_{24}=(-k_y+k_z)/2$, and $\beta_{34}=(k_x-k_y)/2$.
Here $k_x,k_y,k_z$ are reduced Bloch coordinates with the target
pyrochlore lattice's conventional cubic period set to unity.
Eigenvalues are frequency
detunings from the common isolated resonance frequency. The parameters are
signed couplings in the convention of Eq.~(\ref{eq:coupling-overlap}).
For equal lateral and oblique couplings, this is the breathing-pyrochlore
model of Ref.~\cite{Ezawa2018Breathing} with the opposite definition of
the coupling parameters.
The band comparison in Figs.~\ref{fig:fig4}(d,e) retains the high-symmetry
points that become inequivalent in the layered BgV.

\textit{Hyperhoneycomb Hamiltonian.---}
The fcc primitive vectors $g_1,g_2,g_3$ define the basis
$ Q=[g_1,g_2,g_3]$. Mode positions are
$R_{m\mu}=Q(M_{\mathrm{HHC}}m+s_\mu)$:
$m\in\mathbb Z^3$ labels target cells, $\mu=1,\ldots,4$ labels the
basis modes at integer coordinates $s_\mu$ in the fcc primitive basis,
and the columns of $QM_{\mathrm{HHC}}$ are the target translations $A_j$.
The numerical matrices and length scale are given in Supplemental Material,
Note S5 \cite{SupplementalMaterial}.
The embedded coupling matrix $H_{\mathrm{emb}}(\mathcal R;P_{\mathrm{BgV}})$ for the position set
$\mathcal R=\{ R_{ m\mu}\}$, relative to the chosen reference frequency,
must approximate the target Hamiltonian $H_{\mathrm{tar}}$, expressed in
frequency units, using the same basis and boundary conditions.  Errors in
the retained couplings and omitted terms must be small compared with the
retained coupling frequencies and any topological gap.

For the four-site HHC cell in Fig.~\ref{fig:fig4}(f), retaining the
$t_x$, $t_y$, and $t_z$ links gives the Hamiltonian
\begin{equation}
    H_{\mathrm{HHC}}( {k})=
    \begin{pmatrix}
    0 & \eta_x & 0 & \eta_z^{*}\\
    \eta_x^{*} & 0 & t_z & 0\\
    0 & t_z & 0 & \eta_y\\
    \eta_z & 0 & \eta_y^{*} & 0
    \end{pmatrix}.
    \label{eq:hhc}
\end{equation}
Here $\eta_x=t_x(1+e^{-i\theta_1})$,
$\eta_y=t_y(1+e^{-i\theta_2})$, and $\eta_z=t_z e^{i\theta_3}$,
where $\theta_j=k\cdot A_j$ is the Bloch phase under the $j$th
target-cell translation. Eigenvalues are detunings $f-f_0$ from the
reference frequency $f_0=\SI{17.15}{GHz}$.
The model parameters are $t_x=t_y=\SI{-0.051}{GHz}$ and
$t_z=\SI{-0.055}{GHz}$, as extracted from the signed pair couplings
in Supplemental Material, Table S4. The momentum paths are specified
in Note S5 \cite{SupplementalMaterial}.

\end{document}